\documentclass[aps,prb,twocolumn,groupedaddress,floatfix,showpacs,endfloats]{revtex4-1}

\usepackage{graphicx}
\usepackage{amsmath}
\usepackage{subcaption}
\usepackage{color}

\usepackage[OT4]{fontenc}
\usepackage[cp1250]{inputenc}
\newcommand{\parl}{\parallel}

\begin{document}


\title{Electro-optical properties of Cu$_2$O
in the regime of Franz-Keldysh oscillations}


\author{Sylwia Zieli\'{n}ska-Raczy\'{n}ska}
\author{David Ziemkiewicz}
\email{david.ziemkiewicz@utp.edu.pl}
\author{Gerard Czajkowski}
\affiliation{Institute of Mathematics and Physics, UTP University
of Science and Technology, \\ Al. Prof. S. Kaliskiego 7, 85-789
Bydgoszcz, Poland}


\date{\today}

\begin{abstract}
We present the analytical  method enables one to compute the
optical functions i.e., reflectivity, transmission and absorption
including the excitonic effects for a semiconductor  crystal
exposed to a uniform electric field for energy region
 above the gap, for external field suitable to appearance of Franz-Keldysh (FK) oscillations. Our approach intrinsically takes into account  the coherence between the carriers and the
electromagnetic field.
  We quantitatively describe the amplitudes and periodicity of FK modulations and
the influence of Rydberg excitons on FK effect is also taking into account.
Our analytical findings are illustrated numerically for $P$ excitons in Cu$_2$O
crystal.

\end{abstract}


\maketitle

\section{Introduction}

The absorption of light at the fundamental gap in semiconductors
reveals two different kinds of electronic transitions. At photon
energies greater than the energy gap the absorption of light
causes processes in which an electron is transferred to the
conduction band and a hole is left in the valence band. At
energies lower than the gap there are absorption peaks which
correspond to processes in which the conduction band electron and
the valence hole are bound to one another in states within the
forbidden energy gap. The possibility that electrons can exist in
semiconductors in excited bound states, termed excitons, was first
suggested by Frenkel \cite{Frenkel} and Peierls \cite{Peierls}
more than 80 year ago. In analogy to the tight binding and the
nearly free electron approximations, two types of excitons are
considered: tightly bound (Frenkel excitons) and weakly bound
(Wannier-Mott excitons). Frenkel excitons have a Bohr radius of
the order of the lattice constant or smaller. Such exciton is
strongly bound and usually localized on one site. The electron and
the hole do not move independently. In contrast,  the Bohr radius
of a Wannier-Mott exciton is much larger than the lattice
constant, therefore it has a smaller binding energy and is
delocalized over a number of sites. Below we deal with the weakly
bound excitons. In 1952, E. Gross and N. Karriev \cite{Gross}
discovered these Wannier-Mott excitons experimentally in a copper
oxide (Cu$_2$O) semiconductor. After that time excitons  have
remained an important topic of experimental and theoretical
research, because of their  dominant role on the optical
properties of semiconductors (molecular crystals etc.,). The
excitons have been studied in great details in various types of
semiconductor nanostructures and in bulk crystals. Since there is
a inestimable number of papers,  monographs, review articles
devoted to excitons, we refer to only a small collection of them.
\cite{Knox}$^-$\cite{Klingshirn} In 80-ties of the last century
the main stream of research on excitons was focussed on III-V
 and II-VI semiconductors, but the research on the
properties of excitons in Cu$_2$O was continued, in particular, by
the group of Dortmund (\cite{Froelich} and the
references therein).

 A lot of studies, both experimental and
theoretical, have been devoted to examine various properties of
excitons in Cu$_2$O bulk crystal  and it appeared that the
spectroscopical features of copper oxide are through-out
recognized (see Refs.\cite{Froelich}$^-$\cite{Stolz_2012}).
But recently, the interest on this bulk semiconductor
has reborn due to outstanding experiment performed by Kazimierczuk
et al.\cite{Kazimierczuk} who discovered highly excited states,
so-called Rydberg excitons (RE) in the natural crystal of copper
oxide. They have observed absorption lines associated with
excitons of principal quantum numbers up to $n=25$.

A large amount of new studies which focused on extraordinary
properties of RE attracted increasing attention during last three
years (\cite{Thewes}$^-$\cite{Walther}), especially on their
behaviour in external
fields\cite{Zielinska.PRB.2016.b}$^-$\cite{Zielinska.PRB.2016.c}
 as well as in the
context of their spectra similarity to quantum chaos and breaking
of all antiunitary
symmerties\cite{Assmann_symmetry}$^-$\cite{Schweiner_Symmetry}. Recently
the first observation of photoluminescence of excitonic Rydberg states
 has been reported \cite{kitamura}.

Up till now much effort has been devoted to examine excitons in
Cu$_2$O for energies below the gap, the region in which the most
important effect, i.e. the appearance of excitons with high number
$n$, has been observed.
One of the distinction peculiarity of copper oxide is the moderately small Rydberg energy
 of only 90 meV, which assures that all of relevant states from the ground state up to the
 continuum above the band gap are optically accessible using attainable lasers.

 Recently some attempts \cite{Hecktoetter_2017} have been reported , where the optical properties of
 Cu$_2$O for excitation energies exceeding the fundamental gap,
 are examined.
 Heckt\"{o}tter \emph{et al}., \cite{Hecktoetter_2017}
 using  two-color pump-probe spectroscopy, studied RE in copper oxide in the presence of free carriers injected by above-band-gap excitation.
 They examined the impact of an ultra-low-density plasma on  Rydberg excitations at the temperature of a few Kelvin and observed that inside a Cu$_2$O
  crystal plasma shifts the band edge downwards, diminishing the maximum excitable Rydberg state which, in consequence,
  leads to modulation of plasma blockade induced by the band gap
  modulation.

  Below we study an another effect appearing for above-gap
  excitation, which refers to electro-optic properties. For excitation energies below the gap, the main electro-optic effects are the shifting, splitting, and,
  for higher excitonic states, mixing of spectral lines.\cite{Zielinska.PRB.2016.b}
  As it was
  observed for direct-band semiconductors, for energies above the
  gap and when a constant electric field is applied, specific
  oscillations in the spectra has been observed, known as the Franz-Keldysh oscillations.
\cite{Franz}$^-$\cite{Lee}

 These oscillations are results of wave functions "leaking" into the band gap;
the key mechanism of this effect is photon-assisted tunnelling across the bandgap.
When an electric field is applied, the electron and hole wave functions become Airy functions rather than plane waves
(and they have a "tail" which extends into the classically forbidden band gap).
 Due to an electric field influence on interband transitions in the presence of excitons the dielectric constant of a semiconductor exhibits Franz-Keldysh oscillations (FKO),
  which can be detected by modulated reflectance.
Franz-Keldysh effect, which  gives the possibility to create and
control reflectivity oscillations, provide a key ingredient to the
goal of achieving a precise tool for steering on-demand
periodicity and amplitude of electro-modulations. FK effect has
found also practical applications, see, for example,
patents.\cite{Patent}

The theoretical description of the FK effect is quite
different of all phenomena   below the gap. For energies below gap a
well known solution of
a hydrogen-like Schr\"{o}dinger equation can be used, where
the term related to the applied electric field is treated as a
perturbation.\cite{Zielinska.PRB.2016.b} For energies above
the gap one deals with the continuum states. When the electric
field is applied, the relevant material (constitutive)
equation contains terms of different symmetry, so an
analytical solution is not known.

As it was mentioned by Ralph \cite{Ralph} many years ago,  to the best of our  knowledge,
 up till
 now the FK effect was not examined for the
Cu$_2$O bulk crystal.
 In this paper we
propose modification of
 the real density matrix
approach which was applied in our the previous
papers\cite{Zielinska.PRB,Zielinska.PRB.2016.b,Zielinska.PRB.2016.c}
to describe optical properties of Rydberg excitons below gap. Here
we develop this approach including energies above the gap and
taking into account the changes caused by an externally applied
electric field, which intensity should be small enough to avoid
Stark localization but on the other hand sufficient to enable
observation of Franz-Keldysh oscillations. Franz-Keldysh effect
gives the possibility to create and control reflectivity
oscillations. Circumventing this problem would be a key to achieve
the goal to a precise tool for steering on-demand periodicity and
amplitude of electro-modulations.

Below we show that using excitons one gets a flexible tool to study the oscillation dynamics
of reflectivity of a Cu$_2$O crystal  irradiated by an electromagnetic radiation and affected by an electric field.
The tunability, which can be exploited to force the desired period and amplitude of modulations,
can be achieved  through the modification of an external electric field intensity,
which in turn influences the excitonic levels shifting and overlapping.

The paper is organized as follows. In Sec. II we sketch the outline
and present general density matrix equations governing the
evolution of the system, necessary for calculation of the
macroscopic polarization of a medium. This general considerations
are then specified for the case of Cu$_2$O in Sec. III.  In Sec.
IV the electro-susceptibility is studied for the case of
$P$-exciton. Sections  V and VI contain the discussion  how more
excitonic states can be accounted for in the calculations of the
FK effect. The analytical expression for the transmissivity is
presented in Sec. VII and illustrative examples of susceptibility
for Cu$_2$O are examined. The conclusions are discussed in the
last section VIII.

\section{Density matrix formulation}\label{density.matrix}
We intend to calculate the optical functions of a Cu$_2$O crystal,
 when  a homogeneous electric field is applied in the
$z$ direction, which is chosen to be perpendicular to the crystal
surface, and the excitation energy exceeds the fundamental gap
energy. The method is based on the so-called real density matrix
approach (RDMA) which, for a similar physical situation, but with
excitation energy below the gap, was used in
Ref.\cite{Zielinska.PRB.2016.b} The kernel of the RDMA is the
so-called constitutive equation
\begin{eqnarray}\label{constitutiveeqn}
 &&\dot{Y}(\textbf{R},\textbf{r})+({
 i}/\hbar)H_{eh}{Y}(\textbf{R},\textbf{r})+(1/\hbar){\mit\Gamma}{Y}(\textbf{R},\textbf{r})\nonumber\\
 &&=({\rm
 i}/\hbar)\textbf{M}(\textbf{r})\textbf{E}(\textbf{R}),
 \end{eqnarray}
where $Y$ is the bilocal coherent electron-hole amplitude (pair
wave functions), ${\bf R}$ is the excitonic center-of-mass
coordinate, $\textbf{r}=\textbf{r}_e-\textbf{r}_h$ the relative
coordinate, $\textbf{M}(\textbf{r})$ the smeared-out transition
dipole density, ${\bf E}({\bf R})$ is the electric field vector of
the wave propagating in the crystal. The coefficient $\mit\Gamma$
in the constitutive equation  represents dissipative processes.
The two-band effective mass Hamiltonian $H_{eh}$ of the system
under a constant electric field $\textbf{F}=(0,0,F)$ that includes
the electron- and hole kinetic energy terms, the electron-hole
interaction potential and the confinement potentials
\cite{Zielinska.PRB}  has the form

\begin{eqnarray}\label{HehH}
H_{eh}&=&E_{g}-\frac{\hbar^2}{2m_{e}}
\partial^2_{z_e}-\frac{\hbar^2}{2m_{hz}}\partial^2_{z_h}
-\frac{\hbar^2}{2\mu_{\parallel}}
(\partial^2_x+\partial^2_y)\nonumber\\&&-
 \frac{\hbar^2}{2M_{\parallel }}
\left(\partial^2_{R_{\parallel}}+R^{-1}_{\parallel}\partial_{R_{\parallel}}\right)
+
eF(z_h-z_e)\\
&&+V_{eh}(z_e-z_h,\rho)+V_{e}(z_e)+V_{h}(z_h),\nonumber
\end{eqnarray}
\noindent where we have separated the center-of-mass coordinate
$R_{\parallel}$ from the relative coordinate $\rho$ on the plane
$x-y$.

The dipole density vectors $\textbf{M}$ should be chosen
appropriate for \emph{P}- or \emph{F}- excitons.
\cite{Zielinska.PRB.2016.c} The potential term  representing
 the Coulomb interaction in an anisotropic medium is given by

\begin{equation}\label{potential}
V_{eh}=-\frac{e^2}{4\pi\epsilon_0\epsilon_b[(x^2+y^2)+
z^2\epsilon_{\parallel}/\epsilon_z]^{1/2}},
\end{equation}
\noindent where we introduce the two effective dielectric
constants, $\epsilon_{\parallel}$, and $\epsilon_z$, respectively,
and define $\epsilon_b=\sqrt{\epsilon_{\parallel}\epsilon_z}$. The
smeared-out transition dipole density ${\bf M}({\bf r})$, should
be chosen in our case appropriate for \emph{P}- or \emph{F}-
excitons, \cite{Zielinska.PRB.2016.c}
 is related to the bilocality of the amplitude $Y$ and
describes the quantum coherence between the macroscopic
electromagnetic field and the interband transitions.

 The coherent amplitude $Y$ defines
the excitonic counterpart of the polarization
\begin{equation}\label{polarization}
\textbf{P}(\textbf{R})=2 \int {\rm d}^3 r~
\hbox{Re}~\left[\textbf{M}(\textbf{r})Y(\textbf{R},\textbf{r})\right],
\end{equation}
which is than used in the Maxwell propagation equation
\begin{equation}\label{Maxwell}
c^2\nabla_R^2
\textbf{E}-\underline{\underline{\epsilon}}_b\ddot{\textbf{E}}(\textbf{R})=\frac{1}{\epsilon_0}\ddot{\textbf{P}}(\textbf{R}),
\end{equation}
with the use  of the bulk dielectric tensor
$\underline{\underline{\epsilon}}_b$ and the vacuum dielectric
constant $\epsilon_0$. In the present paper we solve the eqs.
(\ref{constitutiveeqn})-(\ref{Maxwell}) in order to compute the
electro-optical functions (i.e.,reflectivity, transmission, and
absorption) for  Cu$_2$O. Contrary to the previous paper on
electro-optical properties \cite{Zielinska.PRB.2016.b} we will
consider the excitation energies above the energy gap, which will
require a different approach.

 Both polarization and electric field must obey Maxwell's equations,
which have to be solved in order to get the propagation modes. The
above approach takes into account key factors necessary for the
calculation of all  optical functions. They are obtained, as
usual, by comparing the amplitudes of incident, reflected or
transmitted electric fields, and depend on the applied field
strength and on the total crystal thickness.
\section{The basic equations}\label{eofunctions}

 The considered crystal  is
modelled by a slab with infinite extension in the $xy$-plane and
the boundary planes $z=0, z=L$. With the sake of simplicity, the
slab is located in vacuum. An monochromatic, linearly polarized
electromagnetic wave propagates along $z$ axis. Its
electric field is given by
\begin{equation}
{\bf E}=(0, E_y, 0),\qquad E_y=E_{in}e^{{ i}k_0z-{ i}\omega t}\,,
\end{equation}
\noindent where $k_0={\omega}/{c}$ is the wave vector in the
vacuum with $\omega$ being the frequency and $E_{in}$ is an
amplitude of the incoming wave.
 Due to the fact that the energy of the propagating wave is
 divided into reflected and transmitted wave one obtains the reflectivity,
transmissivity and absorption   from the relations

\begin{eqnarray}\label{wzoryfcjeoptyczne}
R=\left|\frac{E(0)}{E_{in}}-1\right|^2,&\qquad&
T=\left|\frac{E(z=L)}{E_{in}}\right|^2,\nonumber\\
A&=&1-R-T,
\end{eqnarray}
\noindent where $E(z)$ is the $x$-component of the wave electric
field inside the crystal. The calculation of the optical functions
consists of several steps. The first one is the solution of the
constitutive equation (\ref{constitutiveeqn}). Due to  the
specific properties of Cu$_2$O, we will treat the crystal as the
bulk region and look for the amplitude $Y$ in the form
\begin{equation}
Y(Z,\textbf{r})=Y(\textbf{r})e^{{i}k_zZ}.
\end{equation}
Assuming the wave propagation in the $z$ direction, we neglect the
$R_{\parallel}$ component and arrive at the equation
\begin{eqnarray}\label{Y2}
& &\big[E_{g}-\hbar\omega - {
i}{\mit\Gamma}+\frac{\hbar^2k_z^2}{2M_z}
-\frac{\hbar^2}{2\mu_{z}}\partial^2_{z}
-\frac{\hbar^2}{2\mu_{\parallel}}
\left(\partial^2_x+\partial^2_y\right)
\nonumber\\
& &+eFz+V_{eh}(z,\rho)\big]Y(x,y,z)=\textbf{M}({\bf
r})\textbf{E}(Z),
\end{eqnarray}
where $M_z, \mu_z$ are the exciton total and reduced effective
masses in the $z$-direction, respectively. For further
considerations we must specify the dipole density $M$.
Due to symmetry properties of Cu$_2$O the total symmetry of the excitons' state
must be the same as the symmetry of the dipole operator, and according to this
the transition dipole density appropriate for $P$ exciton will be considered.
 This will result in the shapes of the real and
imaginary part of the electro-susceptibility. In the previous
papers,\cite{Zielinska.PRB.2016.b}$^,$\cite{Zielinska.PRB.2016.c}
for energies below the gap, we took the dipole density in terms of
spherical coordinates. For energies above the gap and with the
applied electric field, the cylindrical symmetry must be used. For
the sake of simplicity, we use the in-plane components of the
dipole density with the coherence radius $r_{0}$  along the planes
and zero in the growth direction. The cylindrical version of the
formulas for $M_x$ and $M_y$ components, given in ref.
\cite{Zielinska.PRB.2016.c},
 has the form
\begin{eqnarray}
\label{emx}{M}_x^{(1)}&\propto&{M}_{0}\sqrt{x^2+y^2} \left(e^{{
i}\phi}-e^{-{
i}\phi}\right)\frac{\exp[-\frac{x^2+y^2}{2r^2_{0}}]}{{
i}\sqrt{\pi} r^3_{0}}\delta(z),\nonumber\\
&&\\ \label{emy} {M}_y^{(1)}&\propto&{M}_{0}\sqrt{x^2+y^2}
\left(e^{{ i}\phi}+e^{-{
i}\phi}\right)\frac{\exp[-\frac{x^2+y^2}{2r^2_{0}}]}{\sqrt{\pi}
r^3_{0}}\delta(z).\nonumber
\end{eqnarray}

For further considerations we introduce dimensionless quantities

\begin{eqnarray}
& &\rho=\frac{\sqrt{x^2+y^2}}{a^*},\quad
\zeta=\frac{z}{a^*{\sqrt{\gamma}}},\\
\nonumber
& &f=\frac{F}{F_{\rm{I}}},\\
\nonumber & &k^2=\frac{2\mu_{\parallel
}}{\hbar^2}a^{*2}\left(E_{g}-{\hbar}{\omega} -{\rm
i}{\mit\Gamma}\right)+\frac{\mu_\parallel}{M_z}\left(k_z^2a^{*2}\right),
\end{eqnarray}
\noindent where $F_{\rm{I}}$ is the so-called ionization field
\begin{equation}
F_{\rm{I}}=\frac{\hbar^2}{2\mu_{\parallel
}ea^{*3}}=\frac{R^*}{a^*e},
\end{equation}
\noindent  with excitonic Rydberg $R^*$, $a^*$ being the
corresponding excitonic Bohr radius, and the anisotropy parameter
$\gamma=\mu_{\parallel H}/\mu_{z}$. With these quantities the Eq.
(\ref{Y2}) can be rewritten in the form

\begin{eqnarray}\label{bulkY}
&&\left(k^2-\partial^2_{\rho}-\frac{1}{\rho}\partial_{\rho}-\frac{1}{\rho^2}\partial^2_\phi
-\partial^2_\zeta+f\sqrt{\gamma}{\zeta}\right )Y\nonumber\\
&&=\frac{2\mu_{\parallel }}{\hbar^2}a^{*2} M^{(1)} E_y+
\frac{2}{\sqrt{\rho^2+{\gamma}{\zeta^2}}}Y.
\end{eqnarray}
$M^{(1)}$  denotes the
relevant components ($x$ or $y$) of the dipole density and  $E_{y}$ is the $y$ component of
 electric wave field.

\section{The electro-susceptibility for the $P$ exciton}

In this section we derive the expression for the bulk Cu$_2$O
electro-susceptibility  for  $P$ exciton. Following the procedure
described in Ref. \cite{Dressler}, we separate the hamiltonian of
Eq. (\ref{constitutiveeqn}) transformed to the form (\ref{bulkY})
into a ``kinetic+electric field'' part $H_{\rm{kin+F}}$
 and a potential term $V$ which lead to the following form of the basic constitutive equations (\ref{constitutiveeqn})
\begin{equation}
H_{\rm{kin}+F}Y={\bf M}{\bf E}-VY.
\end{equation}
\noindent The above expression corresponds to Lippmann-Schwinger
equation, in which the Green function $G$ appropriate to the
``kinetic+electric field'' part is adopted for the coherent
amplitude
\begin{equation}\label{Lippmann}
Y=G{\bf M}{\bf E} - GVY.
\end{equation}
\noindent  The Green function for Eq. (\ref{bulkY}) has the form

\begin{eqnarray}\label{Greenfunction1}
&&G(\rho,\rho';\zeta,\zeta';\phi,\phi')=\frac{1}{(f\sqrt{\gamma})^{\frac{1}{3}}}\\
&& \times\frac{1}{2\pi}\sum\limits_{m=-\infty}^\infty e^{{\rm
i}m(\phi-\phi')}\int\limits_0^{\infty}x\,{\rm
d}x\,J_m(x\rho)J_m(x\rho')g_{x}(\zeta,\zeta'),\nonumber
\end{eqnarray}
\noindent where

\begin{eqnarray}\label{gxzeta}
& &g_{x}(\zeta,\zeta')=g^<\;g^>,\nonumber\\
\nonumber & &g^<=\pi{\rm Bi}\left[(f\sqrt{\gamma})^{\frac{1}{3}}
\left(\zeta^<+\frac{k^2+x^2}{f\sqrt{\gamma}}\right)\right]\nonumber\\
&&+ { i}{\rm Ai}\left[(f\sqrt{\gamma})^{\frac{1}{3}}
\left(\zeta^<+\frac{k^2+x^2}{f\sqrt{\gamma}}\right)\right],\\
\nonumber & &g^>={\rm Ai}\left[(f\sqrt{\gamma})^{\frac{1}{3}}
\left(\zeta^>+\frac{k^2+x^2}{f\sqrt{\gamma}}\right)\right],
\end{eqnarray}
\noindent $J_m$ are Bessel functions, and $\rm{Ai}(z)$,
$\rm{Bi}(z)$ are Airy functions (see Ref.\cite{Abram81}).
\begin{figure}
\includegraphics[width=0.9\linewidth]{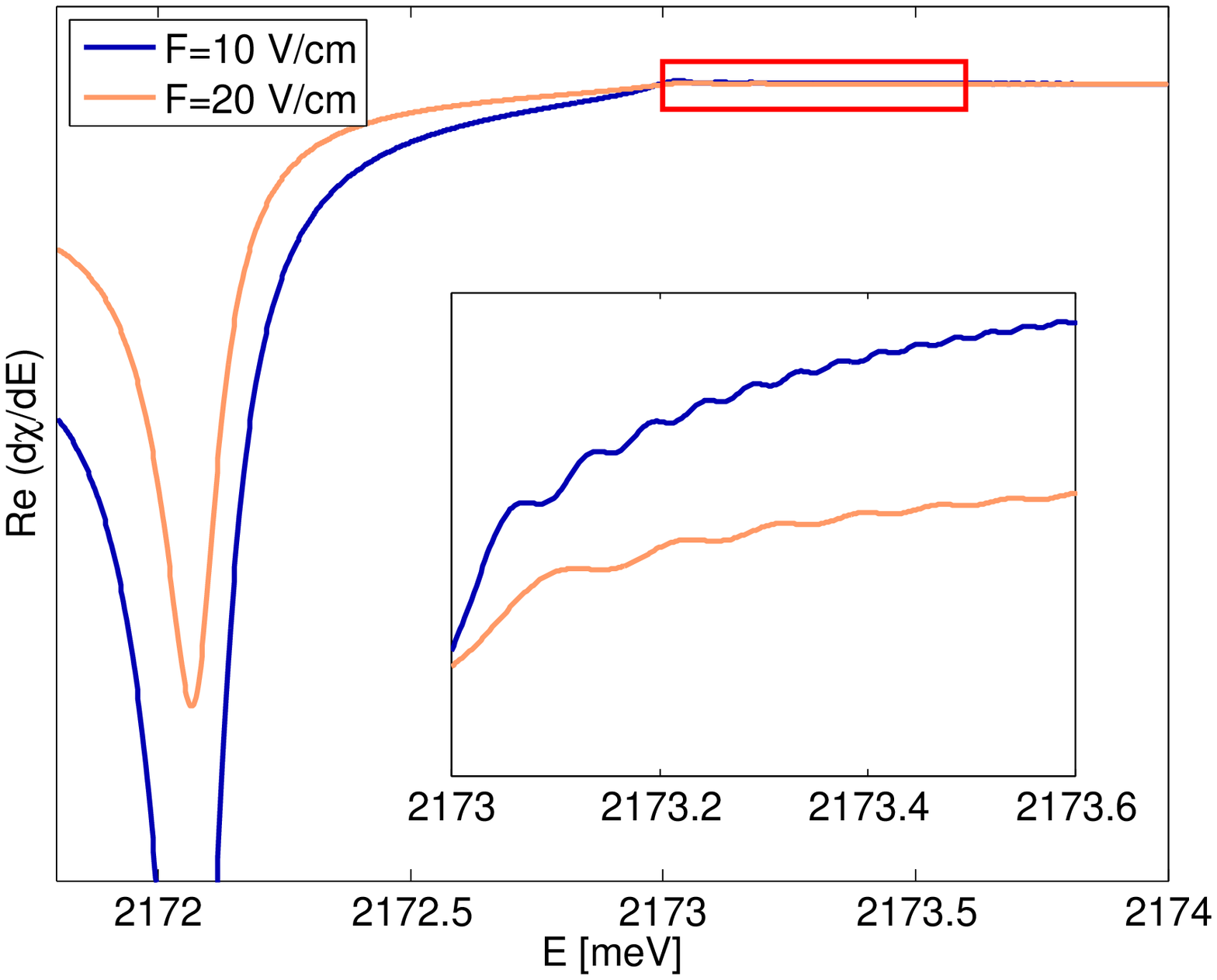}
\caption{\footnotesize Real part of the electro-susceptibility for
a Cu$_2$O crystal, calculated by the formula (\ref{defchi}), taking into account the excitonic effect, for two values
of the electric-field strength (\hbox{$F$=10 V/cm} and \hbox{$F$=20 V/cm}), with the
coherence radius $\rho_0=0.2~a^*$, $\Gamma=0.05$ meV.}\label{Fig1}
\end{figure}
To obtain the optical functions one has to solve the Eq.
(\ref{Lippmann}) using the Green function (\ref{Greenfunction1}).
Please note that the Eq. (\ref{Lippmann}) has the form of Fredholm
integral equation of second type. There are many methods of solving
such equations \cite{Analysis} and particular choice depends on
specific properties of the particular crystal. One of the methods
uses a certain form of the function $Y$ (ansatz) which depends on
an unknown parameter $Y_0$. The parameter is then obtained from
the equation (\ref{Lippmann}) and used to calculate the
polarization from Eq. (\ref{polarization}) and the electric field
of the wave from Eq. (\ref{Maxwell}).

The ansatz for $Y$ will be taken in the form
\begin{equation}
Y(\rho,\zeta,\phi)=Y_{0} \frac{e^{{ i}\phi}+e^{-{
i}\phi}}{2}\,\rho\exp\left(-k\sqrt{\rho^2+\gamma \zeta^2}\right).
\end{equation}
which has the symmetry of 2\emph{P} exciton state.
 With the above ansatz,  Green's function (\ref{Greenfunction1}) and the $M_y$  dipole densities (\ref{emx}),
which in our scaled variables (normalized in spatial variables
$\rho,\zeta$) takes the form
\begin{eqnarray}
M_y^{(1)}&=&\sqrt{\frac{2}{\pi}}\frac{M_{0}}{\pi\sqrt{\gamma}\rho^3_{0}a^{*3}}\nonumber\\
&&\times
\rho\,\exp\left(-\frac{\rho^2}{2\rho^2_{0}}\right)\frac{e^{{\rm
i}\phi}+e^{-{i}\phi}}{2}\delta(\zeta),
\end{eqnarray}
\noindent where $\rho_{0}=r_{0}/a^*$,  we obtain the following
expression for the susceptibility
\begin{figure}
\includegraphics[width=0.9\linewidth]{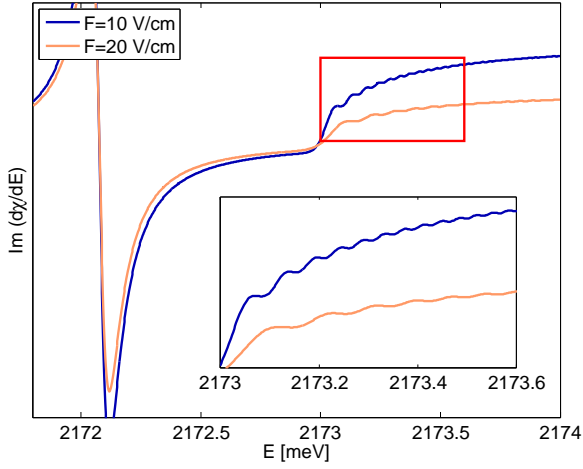}
\caption{\footnotesize The same as in Fig. \ref{Fig1}, for the
imaginary part of the susceptibility}\label{Fig2}
\end{figure}
\begin{eqnarray}\label{defchi}
& &\chi=2\frac{\vert M_{0}\vert^2 2\mu_{\parallel
}\rho_0^2}{\epsilon_0\sqrt{\gamma}a^*\hbar^2
(f\sqrt{\gamma})^{\frac{1}{3}}Q}\nonumber\\
&&\times\int\limits_0^{\infty} x^3 {\rm d}x \exp(-\rho_{0}^2x^2)
\biggl[\mbox{Bi}\left(\frac{k^2_{x}}{(f\sqrt{\gamma})^{\frac{2}{3}}}\right)\\
\nonumber & & +{
i}\mbox{Ai}(\frac{k^2_{x}}{(f\sqrt{\gamma})^{\frac{2}{3}}})\biggr]
\mbox{Ai}\left(\frac{k^2_{x}}{(f\sqrt{\gamma})^{2/3}}\right)
\frac{\tilde{M}Y}{M_{0}Y_{0}}\nonumber\\
&&=2\frac{\vert M_{0}\vert^2 2\mu_{\parallel
}\rho_0^2}{\epsilon_0\sqrt{\gamma}a^*\hbar^2
Q}\cdot\frac{1}{2}(f\sqrt{\gamma})e^{-\mathcal{E}\rho_0^2}\nonumber\\
&&\times\int\limits_{-\mathcal E}^\infty{\rm d}u \,
e^{-\rho_{0}^2(f\sqrt{\gamma})^{2/3}u}\left(u+{\mathcal
E}\right)\left[\mbox{Bi}(u)+i\mbox{Ai}(u)\right]\hbox{Ai}(u),
\nonumber
\end{eqnarray}

\noindent and
\begin{eqnarray}\label{kxH}
& &k^2_{x}=k^2+x^2,\nonumber\\
&&u=\frac{k^2+x^2}{(f\sqrt{\gamma})^{2/3}},\nonumber\\
&&\hbar\Theta=R^*(f\sqrt{\gamma})^{2/3}=\left(\frac{\hbar^2}{2\mu_z}\right)^{1/3}(eF)^{2/3},\\
&&{\mathcal E}=\frac{\hbar\omega-E_g}{\hbar\Theta},\nonumber\\
&&\frac{\tilde{M}Y}{M_{0}Y_{0}}=\frac{\sqrt{2\pi}}{\sqrt{\gamma}}\rho_0\exp\left(\frac{k^2\rho_0^2}{4}\right)D_{-4}(k\rho_0),\nonumber
\end{eqnarray}
$\hbar\Theta$ being the so-called electro-optical energy, and
$D_{-4}$ the parabolic cylinder function (for example,
\cite{{Abram81}}$^,$ \cite{Grad})
\begin{eqnarray}
&&D_p(z)=\frac{\exp(-z^2/4)}{\Gamma(-p)}\int\limits_0^\infty
e^{-xz-(x^2/2)}x^{-p-1}{\rm d}x,\nonumber\\
&&\qquad (\hbox{Re}\,p<0).
\end{eqnarray}
 The expression $Q$ appearing in the
denominator in (\ref{defchi}) is given by

\begin{equation}\label{Q}
Q=\frac{(\tilde{M}Y)-\tilde{M}GVY}{M_0Y_0},
\end{equation}
\noindent where
\begin{eqnarray}\label{MGVY1}
& &\frac{\tilde{M}GVY}{M_0Y_0}=
\frac{2\pi}{(f\sqrt{\gamma})^{1/3}}\int\limits_0^{\infty}{x^3\,{\rm
d}x}
e^{-\rho^2_0x^2/2}\mbox{Ai}\left(\frac{k^2_x}{(f\sqrt{\gamma})^{2/3}}\right)\nonumber\\
 & &\times\int\limits_0^{\infty}{\rm d}\zeta \left(\frac{\sqrt{\gamma}\zeta}{k_x^2}+\frac{1}{k_x^3}\right)e^{-k_x\sqrt{\gamma}\zeta}
\biggl\{ \mbox{Bi}\left[ (f\sqrt{\gamma})^{1/3}\left(\frac{k^2_x}
{f\sqrt{\gamma}}-\zeta\right)\right]\nonumber\\
 & &+ { i}\mbox{Ai}((f\sqrt{\gamma})^{1/3}\left(\frac{k^2_x}
{f\sqrt{\gamma}}-\zeta)\right)\biggr\}\nonumber\\
 & &+\frac{2\pi}{(f\sqrt{\gamma})^{1/3}}
\int\limits_0^{\infty}{x^3\,{\rm d}x e^{-\frac{\rho^2_0x^2}{2}}}
\biggl[\mbox{Bi}\left(\frac{k^2_x}{(f\sqrt{\gamma})^{2/3}}\right)\\
&&+
{i}\mbox{Ai}\left(\frac{k^2_x}{(f\sqrt{\gamma})^{2/3}}\right)\biggr] \nonumber\\
 & &\times \int\limits_0^{\infty}{\rm d}\zeta e^{-k_x
\sqrt{\gamma}\zeta}\left(\frac{\sqrt{\gamma}\zeta}{k_x^2}+\frac{1}{k_x^3}\right)
\mbox{Ai}\left[(f\sqrt{\gamma})^{1/3}\left(\frac{k^2_x}
{f\sqrt{\gamma}}+\zeta\right)\right].\nonumber
\end{eqnarray}

\noindent The vanishing of the real part gives the resonance of
the susceptibility. In particular, for the case without electric field $F=0$, one obtains
\begin{equation}\frac{\tilde{M}GVY}{M_0Y_0}
=\frac{1}{3k}\sqrt{\frac{2}{\pi}}\frac{1}{\sqrt{\gamma}}\frac{2(1+2\sqrt{\gamma})}{(1+\sqrt{\gamma})^2}e^{k^2\rho_0^2/4}\Gamma(4)D_{-4}(k\rho_0),
\end{equation}
with the Euler Gamma function $\Gamma(z)$. Some unique features of
the susceptibility can be read off directly from the formula
(\ref{defchi}). In particular, for energies above the gap (i.e.
for $\hbar\omega>E_g$) we obtain the FK oscillations in the
spectrum, which appear due to periodical character of Airy
functions Ai and Bi.

Some quantitative properties of the spectrum can be obtained by
neglecting the electron-hole interaction, i.e. by taking $V=0$.
The results for the susceptibility ensue from the formula
(\ref{defchi}) with $Q=1$ are shown in Fig.\ref{Fig1} (the real
part) and Fig. \ref{Fig2} (the imaginary part) for two values of
the applied electric field and two values of $\rho_0$. We have
used the values $E_g=2172~\hbox{meV}, R^*=86.981~\hbox{meV}$,
$\gamma=0.5351$, $\mu_{\parallel}=0.396\;m_0, a^*=1.0\;
\hbox{nm}$, and phenomenological value of damping
${\mit\Gamma}=0.05~\hbox{meV}$. The dipole matrix
element $M_0$ is related to the longitudinal-transverse splitting
energy $\Delta_{LT}$.\cite{Zielinska.PRB}

\begin{figure}
\includegraphics[width=0.9\linewidth]{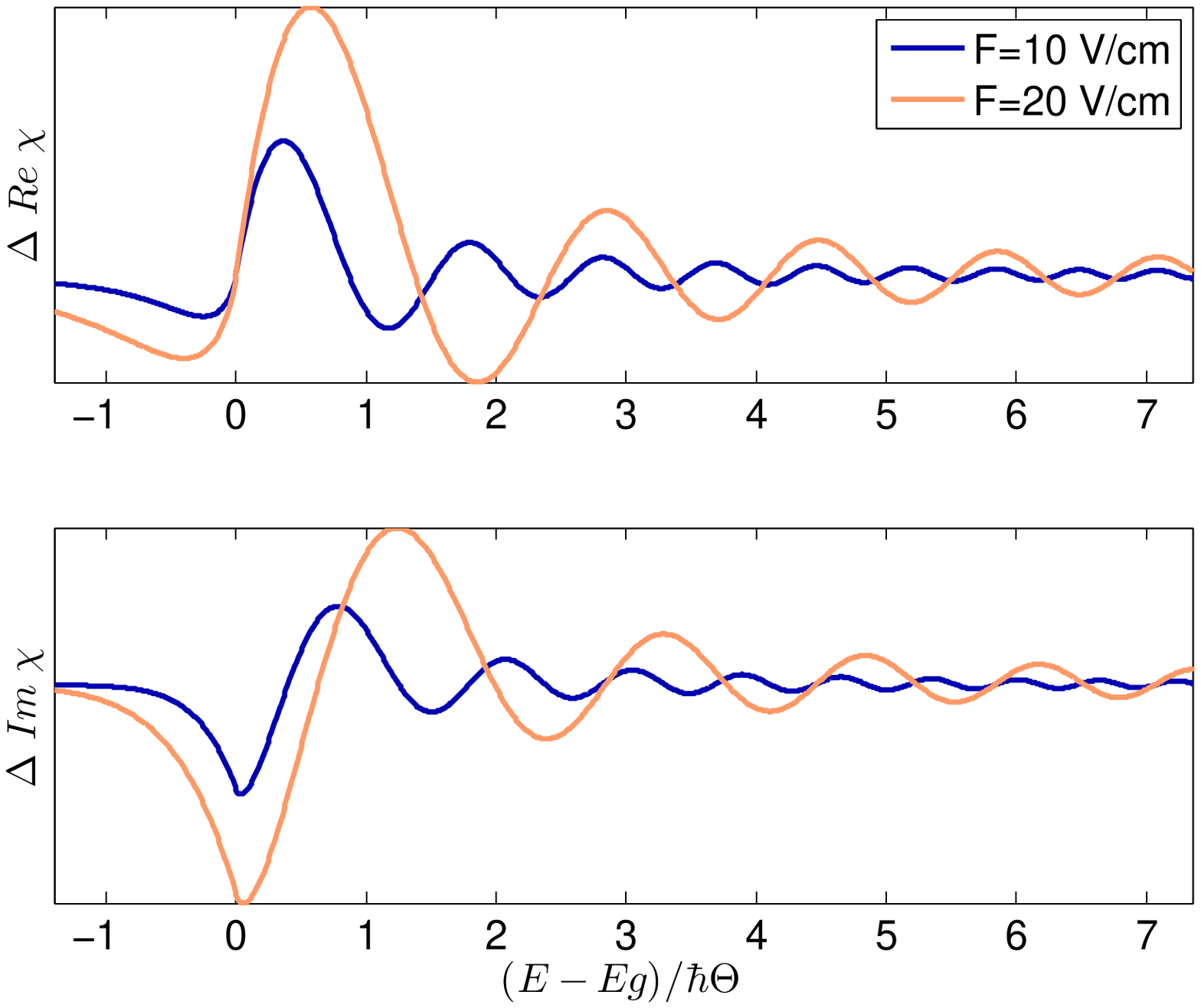}
\caption{\footnotesize The difference $\Delta
\chi=\chi(F)-\chi(F=0)$ displayed
for the data of Fig. \ref{Fig1}, for two values of the applied
electric field.}\label{Fig3}
\end{figure}

It can be seen from Fig.\ref{Fig1}-\ref{Fig2}
 that for energies above the gap  the noticeable oscillations
in dispersion and absorption spectra appear.
Their period and amplitude increase with a fields strength.
It should be noted that the external field should be chosen carefully, i.e.,
to be small enough to avoid Stark localization but sufficiently strong for
oscillation to manifest.
The results of the field impact is more pronounced on susceptibility differential spectrum
\begin{equation}
\Delta\chi=\chi(F)-\chi(F=0).
\end{equation}
 Fig. \ref{Fig3} presents the imaginary part of the
susceptibility for two values of the external electric fields.
One can see more explicit the oscillations of real and imaginary part of $\chi$;
their amplitudes are slightly dependent on the the field strength while oscillation period
strongly depends on the field; this relation will be discussed in more details below.

The results are more evident when we consider the limit
${\mit\Gamma}\to 0, k_z\to 0, \rho_0\to 0$, which is suitable for
situation above the gap. Then the integrals in
(\ref{defchi}) involving Airy functions can be
performed\cite{Airy} and we obtain
\begin{eqnarray}\label{Rechirhoeqzero}
&&\mbox{Re}\,\chi=\frac{4\chi'\sqrt{\pi}\rho_0^3}{Q}\left(\frac{\hbar\Theta}{R^*}\right)^{3/2}\nonumber\\
&&\times\frac{1}{6}\left[4{\mathcal
E}\hbox{Ai}'\hbox{Bi}'+4{\mathcal
E}^2\hbox{Ai}\hbox{Bi}-\hbox{Ai}'\hbox{Bi}-\hbox{Ai}\hbox{Bi}'\right]\\
&&=\frac{4\chi'\sqrt{\pi}\rho_0^3}{Q}\biggl\{\frac{2}{3}\left(\frac{\hbar\Theta}{R^*}\right)^{1/2}\nonumber\\
&&\times\left(\frac{\hbar\omega-E_g}{R^*}\right)\hbox{Ai}'\left(-\frac{\hbar\omega-E_g}{\hbar\Theta}\right)\hbox{Bi}'
\left(-\frac{\hbar\omega-E_g}{\hbar\Theta}\right)\nonumber\\
&&+\frac{2}{3}\left(\frac{R^*}{\hbar\Theta}\right)^{1/2}\left(\frac{\hbar\omega-E_g}{R^*}\right)^2\nonumber\\
&&\times
\hbox{Ai}\left(-\frac{\hbar\omega-E_g}{\hbar\Theta}\right)\hbox{Bi}\left(-\frac{\hbar\omega-E_g}{\hbar\Theta}\right)\nonumber\\
&&-\frac{1}{6}\left(\frac{\hbar\Theta}{R^*}\right)^{3/2}\hbox{Ai}'\left(-\frac{\hbar\omega-E_g}{\hbar\Theta}\right)\hbox{Bi}
\left(-\frac{\hbar\omega-E_g}{\hbar\Theta}\right)\nonumber\\
&&-\frac{1}{6}\left(\frac{\hbar\Theta}{R^*}\right)^{3/2}\hbox{Ai}\left(-\frac{\hbar\omega-E_g}{\hbar\Theta}\right)\hbox{Bi}'
\left(-\frac{\hbar\omega-E_g}{\hbar\Theta}\right)\biggr\}\nonumber\\
&&\nonumber
\end{eqnarray}
\begin{eqnarray}\label{Imchirhoeqzero}
&&\hbox{Im}\,\chi=-C\biggl\{\frac{1}{6}\left(\frac{\hbar\Theta}{R^*}\right)^{3/2}\hbox{Ai}'\left(-\frac{\hbar\omega-E_g}{\hbar\Theta}\right)\hbox{Ai}
\left(-\frac{\hbar\omega-E_g}{\hbar\Theta}\right)\nonumber\\
&&+\frac{1}{3}\left(\frac{\hbar\Theta}{R^*}\right)^{1/2}
\left(\frac{\hbar\omega-E_g}{R^*}\right)\hbox{Ai}'^2\left(-\frac{\hbar\omega-E_g}{\hbar\Theta}\right)\nonumber\\
&&+\frac{1}{3}\left(\frac{R^*}{\hbar\Theta}\right)^{1/2}\left(\frac{\hbar\omega-E_g}{R^*}\right)^2
\hbox{Ai}^2\left(-\frac{\hbar\omega-E_g}{\hbar\Theta}\right)\biggr\}
\end{eqnarray}
with a certain constant C, and $\chi'$ defined as
$$\chi'=\frac{\epsilon_b\sqrt{\pi}\Delta_{LT}}{8R^*\sqrt{\gamma}{\rho_0}}.$$

 Note that the
above expressions differ from the expressions for
\emph{S}-excitons (allowed interband transitions, for example
GaAs)\cite{Tharmalingam,callaway,Dressler}
\begin{eqnarray}
&&\hbox{Im}\,\chi\propto
\biggl\{\frac{\hbar\omega-E_g}{\hbar\Theta}\hbox{Ai}^2\left(-\frac{\hbar\omega-E_g}{\hbar\Theta}\right)\nonumber\\
&&+\left[\hbox{Ai}'
\left(-\frac{\hbar\omega-E_g}{\hbar\Theta}\right)\right]^2\biggr\}.\end{eqnarray}

Now we can perform qualitative discussion  of the spectra spectra features.
Having in mind the properties of Cu$_2$O we observe, that the
value of the electro-optical energy $\hbar\Theta$ is small compared
to the Rydberg energy. Therefore the arguments of the Airy
functions in expressions (\ref{Rechirhoeqzero}) and
(\ref{Imchirhoeqzero}) quickly reach the values which justify the
use of their asymptotic expansions, giving in the lowest order with respect to $\zeta$
the formulas for the real and imaginary part of the
susceptibility
\begin{eqnarray}\label{asymptotic}
&&\hbox{Re}\,\chi\to
C_1\left(\frac{\hbar\Theta}{R^*}\right)^{3/2}\sin
2\zeta,\\
&&\hbox{Im}\,\chi\to-C_2\left(\frac{\hbar\Theta}{R^*}\right)^{3/2}\cos
2\zeta+\frac{1}{3\pi}\left(\frac{\hbar\omega-E_g}{R^*}\right)^{3/2}\nonumber\\
&&\zeta=\frac{2}{3}\left(\frac{\hbar\omega-E_g}{\hbar\Theta}\right)^{3/2},\nonumber
\end{eqnarray}
with certain constants $C_1,C_2$. The above expressions allow to  get the
periodicity of the FK oscillations; the
peaks will appear at energies
\begin{equation}\label{FKE}
(E_n-E_g)^{{3}/{2}}=\frac{3}{4}n\pi (\hbar\theta)^{3/2}=
\frac{3n{\pi}e\hbar F}{4\sqrt{2\mu_z}}.
\end{equation}
\noindent Please note that the above formula includes all extrema.
 It means that the periodicity of FK oscillations
can be used for determining the effective masses along the $z$
axis, as it was done in the case of semiconductor
superlattices.\cite{Schlichterle,Nakayama} With regard to Eq.
(\ref{asymptotic}) we observe FK oscillations around a curve
$(\hbar\omega-E_g)^{3/2}$. The slope is analogous to that obtained
for forbidden transitions\cite{Tharmalingam,callaway,Schaevitz} and differs
from that observed for $S$ excitons, which in turn depend on
$({\hbar\omega-E_g})^{1/2}$.\cite{Tharmalingam,callaway,Dressler}

\section{Impact of higher excitonic states on Franz-Keldysh effect}\label{Appendix B}

Above we have considered the Franz-Keldysh effect with
 one  exciton state.
 Up till now  only the problem of the dependence of  multiplicity of excitonic states on Franz-Keldysh effect
 for confined  systems or for a system in an external magnetic field were examined (\cite{Festschrift,RivistaGC} and references therein),  but the general solution of the issue for bulk crystal is not available.  Below we propose a
method which allows to study the effect of two lowest exciton
states. To achieve this goal  we will consider the amplitude $Y$ in the form
\begin{equation}\label{Ytwostates}
Y=Y_1+Y_2,
\end{equation}
where
\begin{eqnarray}
&&Y_1=N_1Y_{01}\rho
e^{i\phi}e^{-k\sqrt{\rho^2+\gamma^2\zeta^2}}=Y_{01}\psi_1,\nonumber\\
&&Y_2=N_2Y_{02}\rho
e^{i\phi}\left(3-k\sqrt{\rho^2+\gamma^2\zeta^2}\right)e^{-(2k/3)\sqrt{\rho^2+\gamma^2\zeta^2}}\nonumber\\
&&=Y_{02}\psi_2,
\end{eqnarray}
orthogonal for $k$ real (below the gap for $\Gamma$=0), and $N_1,
N_2$ represent the normalization factors for the resonance
energies. In the above definitions we neglect the center-of-mass
dependence.

 \noindent The $ansatz$ (\ref{Ytwostates}) contains two unknown
 parameters $Y_{01},Y_{02}$. They can be determined from the integral
 equation (\ref{Lippmann}). One of the possible methods is to use
 the projection of those equations onto an orthonormal basis which yields equations for the parameters (Galerkin method).
 Wee choose the basis in the form
\begin{eqnarray}
\varphi_1(\rho,\zeta,\phi)&=&\frac{1}{\sqrt{\pi}}\frac{\rho}{\rho_0^2} e^{i\phi}\exp\left(-\frac{\rho^2}{2\rho_0^2}\right)\delta(\zeta),\\
\varphi_2(\rho,\zeta,\phi)&=&\sqrt{\frac{2}{\pi}}\frac{\rho}{\rho_0^2}e^{i\phi}\left(1-\frac{\rho^2}{2\rho_0^2}\right)\exp\left(-\frac{\rho^2}{2\rho_0^2}\right)\delta(\zeta).\nonumber
\end{eqnarray}
\noindent Using the common notation for scalar product we obtain two equations
\begin{eqnarray}\label{Galerkin}
\langle\varphi_1\vert Y\rangle&=&\langle\varphi_1\vert GM\rangle-\langle\varphi_1\vert GVY\rangle,\nonumber\\
\langle\varphi_2\vert Y\rangle&=&\langle\varphi_2\vert
GM\rangle-\langle\varphi_2\vert GVY\rangle,
\end{eqnarray}
\noindent where we neglected the constant factors. Inserting the
expression for $Y$ we get from (\ref{Galerkin}) the equations
\begin{eqnarray}\label{uklad}
a_{11}x_1+a_{12}x_2&=&b_1,\nonumber\\
a_{21}x_1+a_{22}x_2&=&b_2,
\end{eqnarray}
\noindent where
\begin{eqnarray}
x_1&=& \frac{2}{\epsilon_0{\mathcal E}}M_{0}Y_{01}, \quad x_2~=~\frac{2}{\epsilon_0{\mathcal E}}M_{0}Y_{02}\nonumber,\\
a_{11}&=&\langle\varphi_{1}\vert
\psi_{1}\rangle-\langle\varphi_{1}\vert
G\tilde{V}\psi_{1}\rangle,\end{eqnarray} \noindent and
\begin{eqnarray}
a_{12}&=&\langle\varphi_{1}\vert\psi_{2}\rangle-\langle\varphi_{1}\vert G\tilde{V}\psi_{2}\rangle\nonumber,\\
a_{21}&=&\langle\varphi_{2}\vert\psi_{1}\rangle-\langle\varphi_{2}\vert G\tilde{V}\psi_{1}\rangle,\nonumber\\
a_{22}&=&\langle\varphi_{2}|\psi_{2}\rangle-\langle\varphi_{2}\vert
G\tilde{V}\psi_{2}\rangle,\nonumber\\
&&\\
\tilde{V}&=&\frac{2}{\sqrt{\rho^2+\gamma\zeta^2}},\qquad \tilde{M}=\frac{M}{M_{0}},\nonumber\\
b_1&=&\frac{2\mu_{\parl}}{\hbar^2 a^*}\frac{2M_{0}^2}{\epsilon_0}\langle\varphi_{1}\vert G\tilde{M}\rangle,\nonumber\\
b_2&=&\frac{2\mu_{\parl}}{\hbar^2
a^*}\frac{2M_{0}^2}{\epsilon_0}\langle\varphi_{2}\vert
G\tilde{M}\rangle.\nonumber
\end{eqnarray}

\noindent The quantities $x_1,x_2, b_1, b_2$ are dimensionless;
$x_1,x_2$ define the electro-susceptibility by the equation
\begin{equation}
\chi=x_1\langle M\vert\psi_1\rangle+x_2\langle
M\vert\psi_2\rangle.
\end{equation}
As it has been done above, some information can be elicit by setting
$V=0$. After a simple algebra we obtain
\begin{eqnarray}\label{x1x2}
&&x_1=\frac{1}{\Delta}\left(b_1\langle\varphi_2\vert\psi_2\rangle-b_2\langle\varphi_1\vert\psi_2\rangle\right),\nonumber\\
&&x_2=\frac{1}{\Delta}\left(b_2\langle\varphi_1\vert\psi_1\rangle-b_1\langle\varphi_2\vert\psi_1\rangle\right),\\
&&\Delta=\langle\varphi_1\vert\psi_1\rangle\langle\varphi_2\vert\psi_2\rangle-\langle\varphi_1\vert\psi_2\rangle\langle\varphi_2\vert\psi_1\rangle,\nonumber
\end{eqnarray}

Comparing the above outcomes with the aforementioned results for
one exciton state we observe that an additional state (the same
holds for more additional states) will only influence the shape and the
amplitude of FK oscillations while the periodicity will practically
remain the same since it is involved in the Green function. One can
also say that the calculation, at least the analytical one,
will be more intricate than in the case of the one exciton state. In
consequence, the method described above practically is operational
only for two exciton states. However, it should be also stressed that the higher
excitonic states are coupled with oscillator strengths decreasing as
$1/n^3$, so their influence will at least be
orders of magnitude smaller that contribution of the two
lowest states.
\section{Rydberg excitons in a one-dimensional model}
As we have discussed above, the simultaneous description of a
multiplicity of excitonic states below the gap and the FK
oscillations above the gap is, at the moment, not accessible. So,
 considering the multiplicity of exciton states as the
dominant feature of Rydberg excitons, we propose a simplified
exciton model, where both phenomena can be described by analytical
formulas. To this end, we consider a system with the reduced
dimensionality, where the electron with the effective mass $m_e$
and the hole with the effective mass $m_{hz}$ move along the
$z$-axis. A constant electric field $F$ is applied in the same
direction. The optical properties of the system described in
previous sections will be described with the RDMA, starting from
the constitutive equation (\ref{constitutiveeqn}), with the
Hamiltonian (\ref{HehH}) which now takes the form
\begin{eqnarray}\label{HehH1dim}
H_{eh}&=&E_{g}-\frac{\hbar^2}{2m_{e}}
\partial^2_{z_e}-\frac{\hbar^2}{2m_{hz}}\partial^2_{z_h}
+
eF(z_h-z_e)\nonumber\\
&&+V_{eh}(z_e-z_h)+V_{e}(z_e)+V_{h}(z_h).
\end{eqnarray}
To account for $n$ excitonic states, we consider the system as a
set of independent oscillators which, in our formalism, will be
related to the exciton amplitudes $Y_n$. The amplitudes will
satisfy the equations
\begin{eqnarray} 
\label{Y21} & & \Biggl[ E_{g}-\hbar\omega- i\Gamma_{n}-
\frac{\hbar^2}{2\mu_{z}}\partial^2_{z} \nonumber\\ & &{} +
eFz+V_{ehn}(z)\Biggr]Y_{n}(z)= M_{n}(z) E_0,
\end{eqnarray} 
\noindent where $E_0$ is the amplitude of the electromagnetic wave
propagating in the medium. The potentials $V_{ehn}$ and the
transition dipole matrix elements $M_n$ will be chosen to
reproduce the optical properties of Rydberg excitons. The Eq.
(\ref{polarization}) for the total polarization will be replaced
by the relation
\begin{equation} 
\label{PR} {\bf P}({\bf R})=2\int {\rm d}^3{ r}\,\mbox{Re}\left[
\sum_{n} M_{n}({\bf r})Y_{n}({\bf r},{\bf R})\right].
\end{equation} 
\noindent

\begin{figure}
\includegraphics[width=0.9\linewidth]{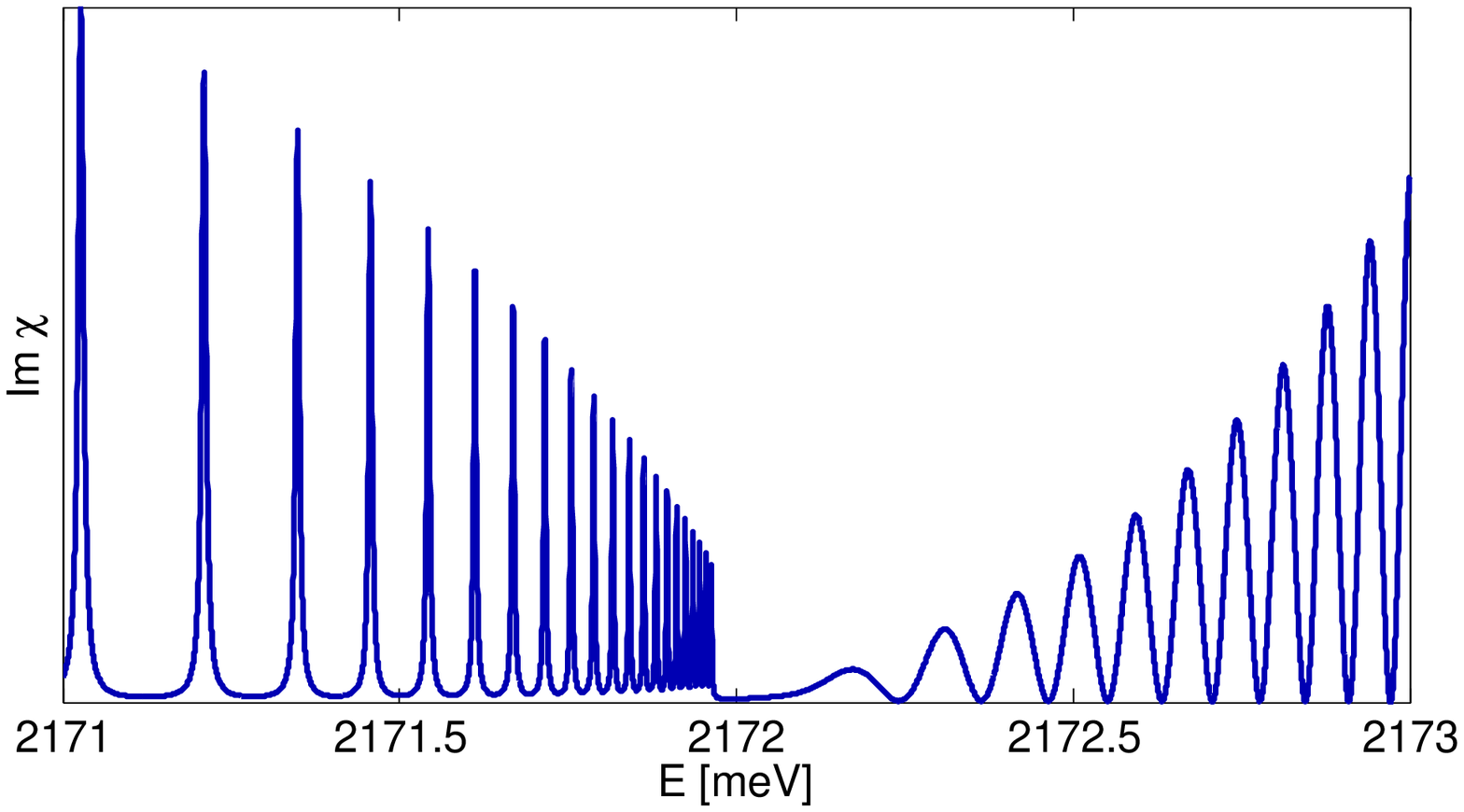}
\caption{\footnotesize Aborption spectrum in a one-dimensional
model, calculated by the formula (\ref{chifield}), taking into
account the excitonic states with n=2-20.}\label{Fig_one1}
\end{figure}

\begin{figure}
\includegraphics[width=0.9\linewidth]{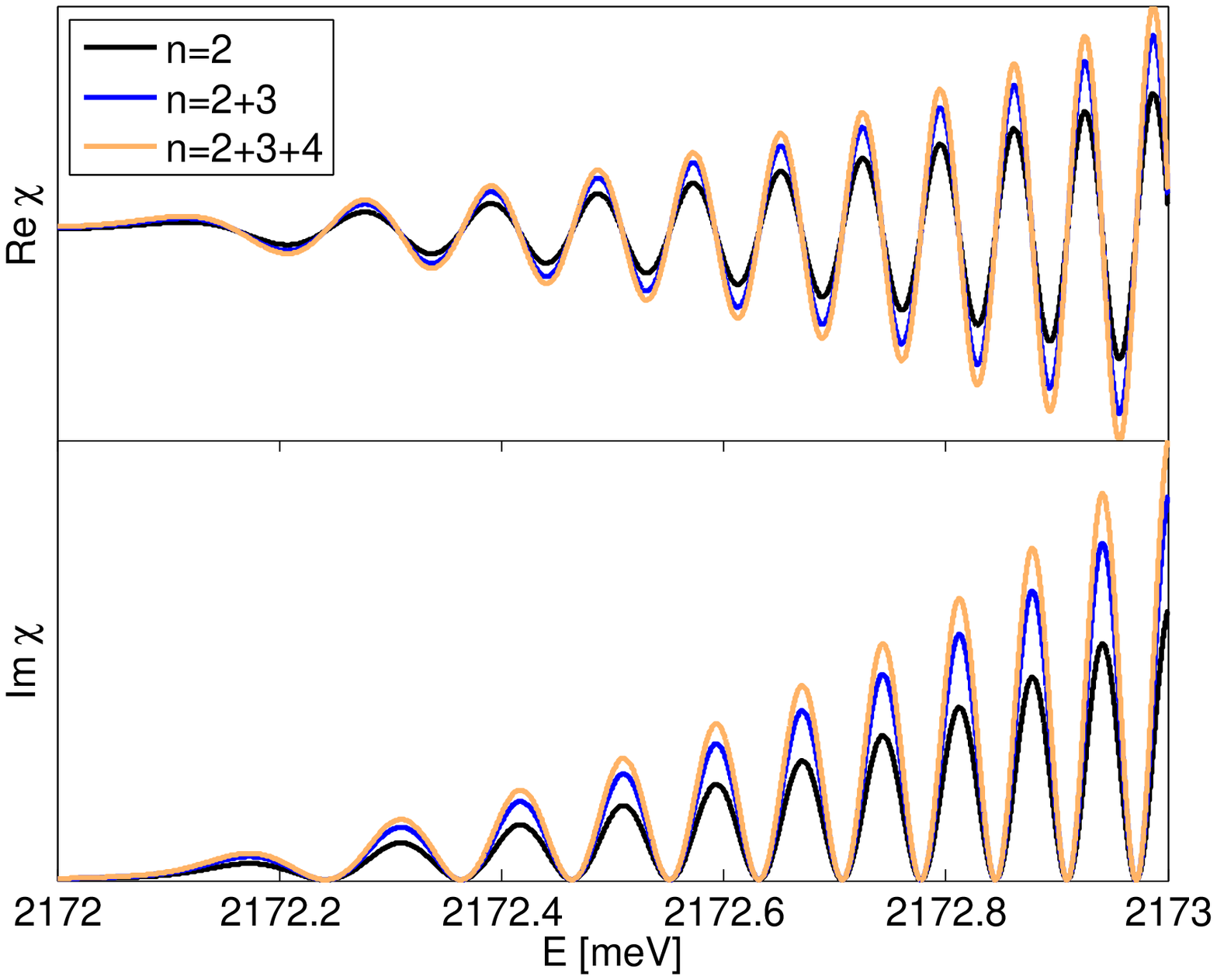}
\caption{\footnotesize Real and imaginary part of susceptibility
in the energetic region above the gap.}\label{Fig_one2}
\end{figure}

Following the scheme described in Sec. III, we arrive at the
equation
\begin{eqnarray}\label{1dimY}
&&\left[k^2-\partial_\zeta^2+f\sqrt{\gamma}\zeta\right]Y_n\nonumber\\
&&=\frac{2\mu_{\parallel}}{\hbar^2}a^{*2}\tilde{M}_n(\zeta)E_0-\tilde{V}_{ehn}(\zeta)Y_n.\end{eqnarray}
The Green function of the above equation has the form (compare Eq.
(\ref{gxzeta}) )
\begin{eqnarray}\label{gzeta1}
& &G(\zeta,\zeta')=g^<\;g^>,\nonumber\\
\nonumber & &g^<=\pi{\rm Bi}\left[(f\sqrt{\gamma})^{\frac{1}{3}}
\left(\zeta^<+\frac{k^2}{f\sqrt{\gamma}}\right)\right]\nonumber\\
&&+ { i}{\rm Ai}\left[(f\sqrt{\gamma})^{\frac{1}{3}}
\left(\zeta^<+\frac{k^2}{f\sqrt{\gamma}}\right)\right],\\
\nonumber & &g^>={\rm Ai}\left[(f\sqrt{\gamma})^{\frac{1}{3}}
\left(\zeta^>+\frac{k^2}{f\sqrt{\gamma}}\right)\right].
\end{eqnarray}
When the external electric field is absent, the Green function
takes the form
\begin{equation}\label{GF0}
G(\zeta,\zeta')_{F=0}=\frac{\exp(-k\vert \zeta-\zeta'\vert}{2k}.
\end{equation}
Choosing $\tilde{M}_n$ and $\tilde{V}_{ehn}$ in the form
\begin{eqnarray}
&&\tilde{M}_n=M_{0n}\delta(\zeta),\quad
\tilde{V}_{ehn}=2\sqrt{\varepsilon_{Tn}}\delta(\zeta)
\end{eqnarray}
we arrive at the following expression for the susceptibility
\begin{equation}\label{chifield}
\chi=\sum\limits_n \frac{f_n
G(0,0)}{1-2\sqrt{\varepsilon_{Tn}}G(0,0)},
\end{equation}
with oscillator strength, for which we can use the expressions
derived in Ref.\cite{Zielinska.PRB} With respect to (\ref{GF0}),
for energies below the gap and for the field $F=0$, we obtain
\begin{equation}\label{chi1dim}
\chi=\sum\limits_n\frac{f_n}{2(k-\sqrt{\varepsilon_{Tn}})}.
\end{equation}
The poles in the susceptibility  define the quantities
$\varepsilon_{Tn}$ which can be expressed by the exciton
resonances energies $\hbar\omega_{Tn}$ as
$\varepsilon_{Tn}=(E_g-\hbar\omega_{Tn})/R^*$. When considering
the case of Cu$_2$O, the resonance energies are well-known, both
experimentally,\cite{Kazimierczuk}$\,$as
theoretically.\cite{Zielinska.PRB,Schweiner_2017a} For Cu$_2$O we
start with $n=2$ and the oscillator strengths $f_n$ will be chosen
as
\begin{equation}
 f_n=\epsilon_b\frac{\Delta_{LT}}{R^*}\frac{32}{3}\frac{n^2-1}{n^5}
.\end{equation} The absorption calculated by the Eq.
(\ref{chi1dim}) is shown in  Fig. \ref{Fig_one1}. One can see the
resonances below the gap as well as characteristic for FK effect
oscillations in the region above the gap. The Rydberg excitones
states compose the background of FK oscillations. When the
electric field is applied, we use the expression (\ref{chifield})
using the Green function (\ref{gzeta1}). As in the 3-dimensional
considerations, we observe the FK oscillations (Fig.
\ref{Fig_one2}). One can see that the period and phase of the them
do not depend on excitonic state number \emph{n}. The advantage of
the method described in this section results from the fact that
arbitrary number of excitonic states can be taken into account; in
such a case the optical functions display the impact of the
increasing number of states taken into account.

\section{Optical functions and exciton effect}
\begin{figure}
a)\includegraphics[width=0.8\linewidth]{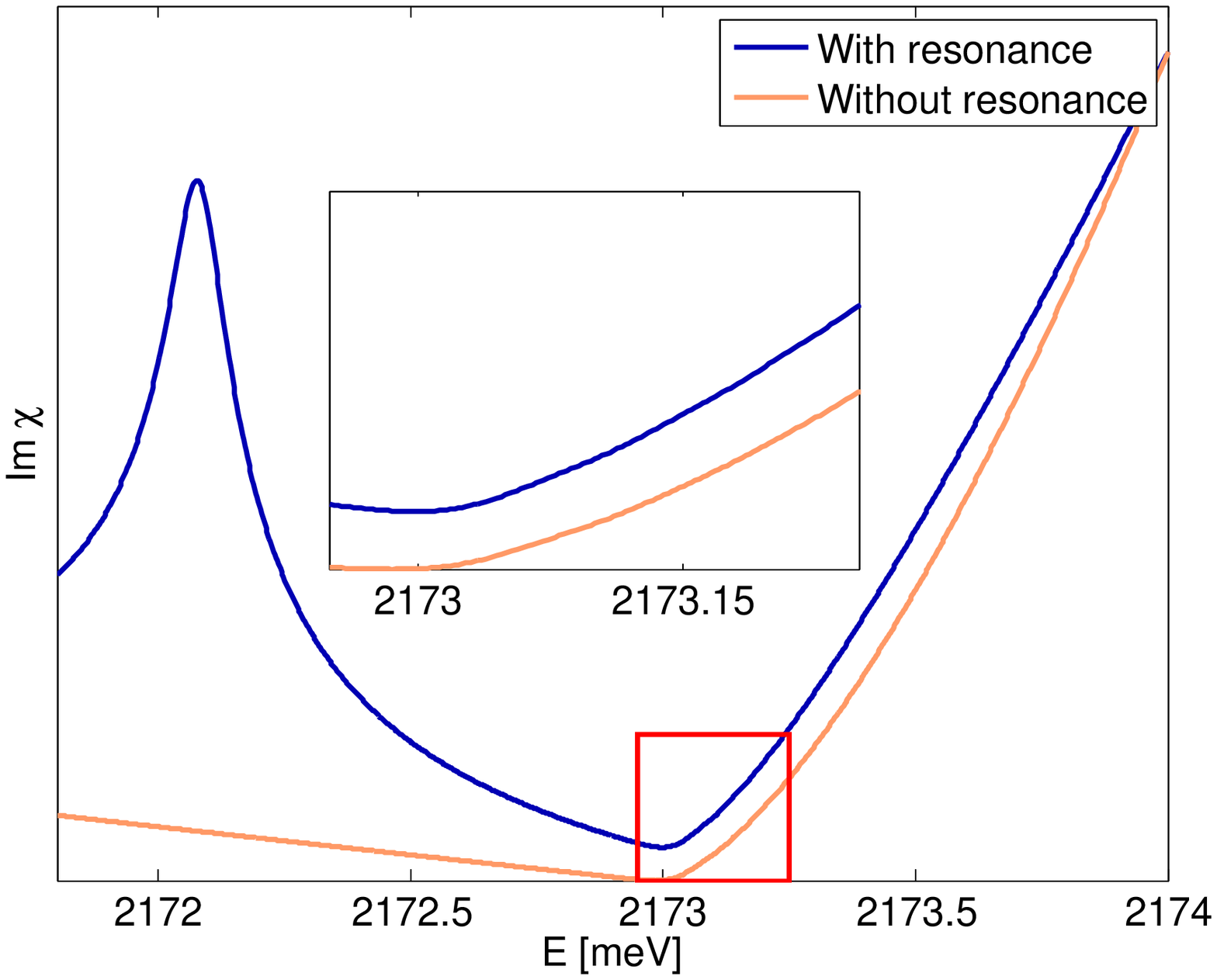}
b)\includegraphics[width=0.8\linewidth]{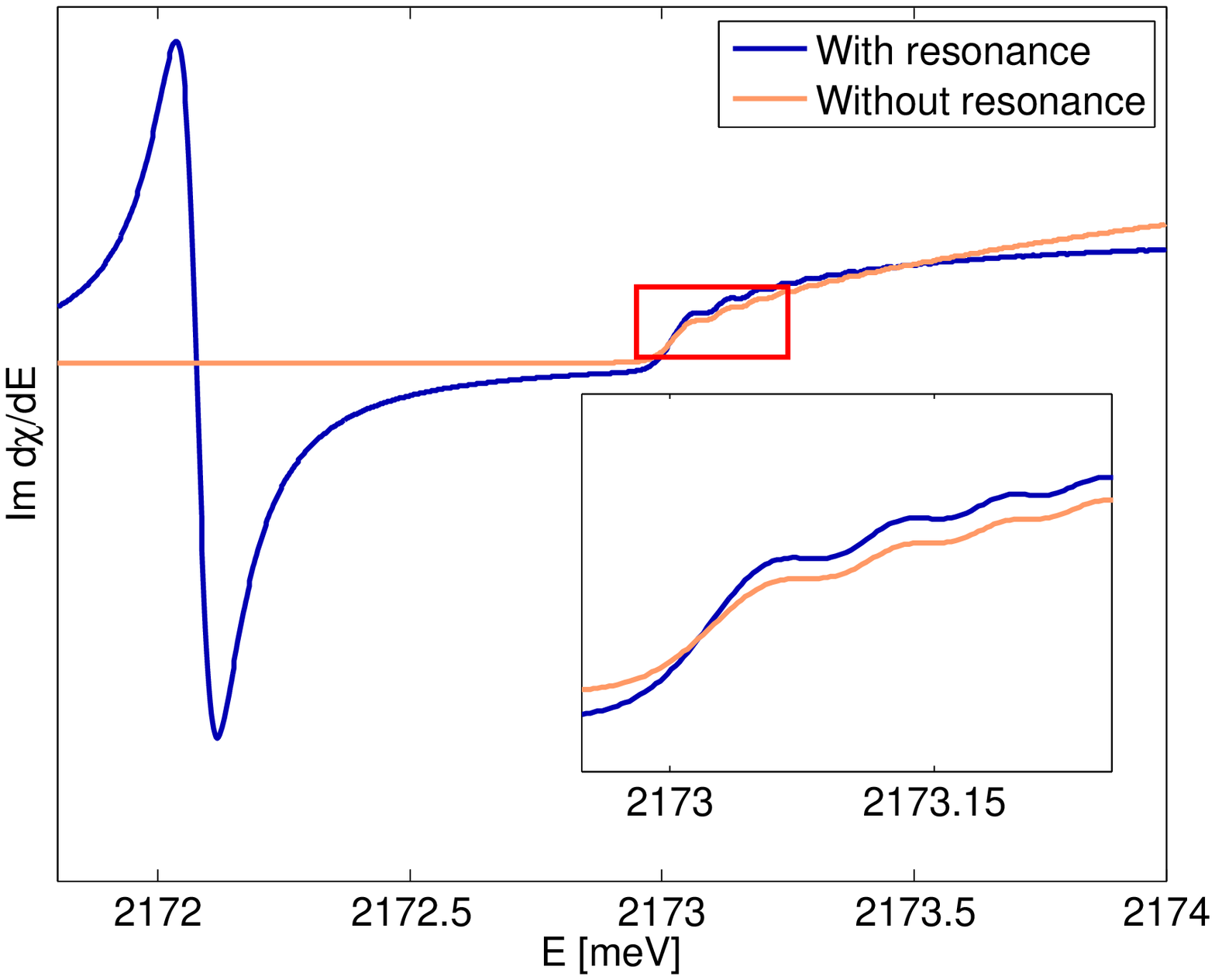}
\caption{\footnotesize (a) Imaginary part of the susceptibility
for a Cu$_2$O crystal, calculated by Eq. (\ref{defchi}), by taking
into account one exciton state and with the applied field strength
\hbox{F=10 V/cm}, compared with those calculated without
excitonic effects, with parameters as in Fig. \ref{Fig2}, (b) The first derivative of susceptibility.}\label{Fig4}
\end{figure}

\begin{figure}
\includegraphics[width=0.9\linewidth]{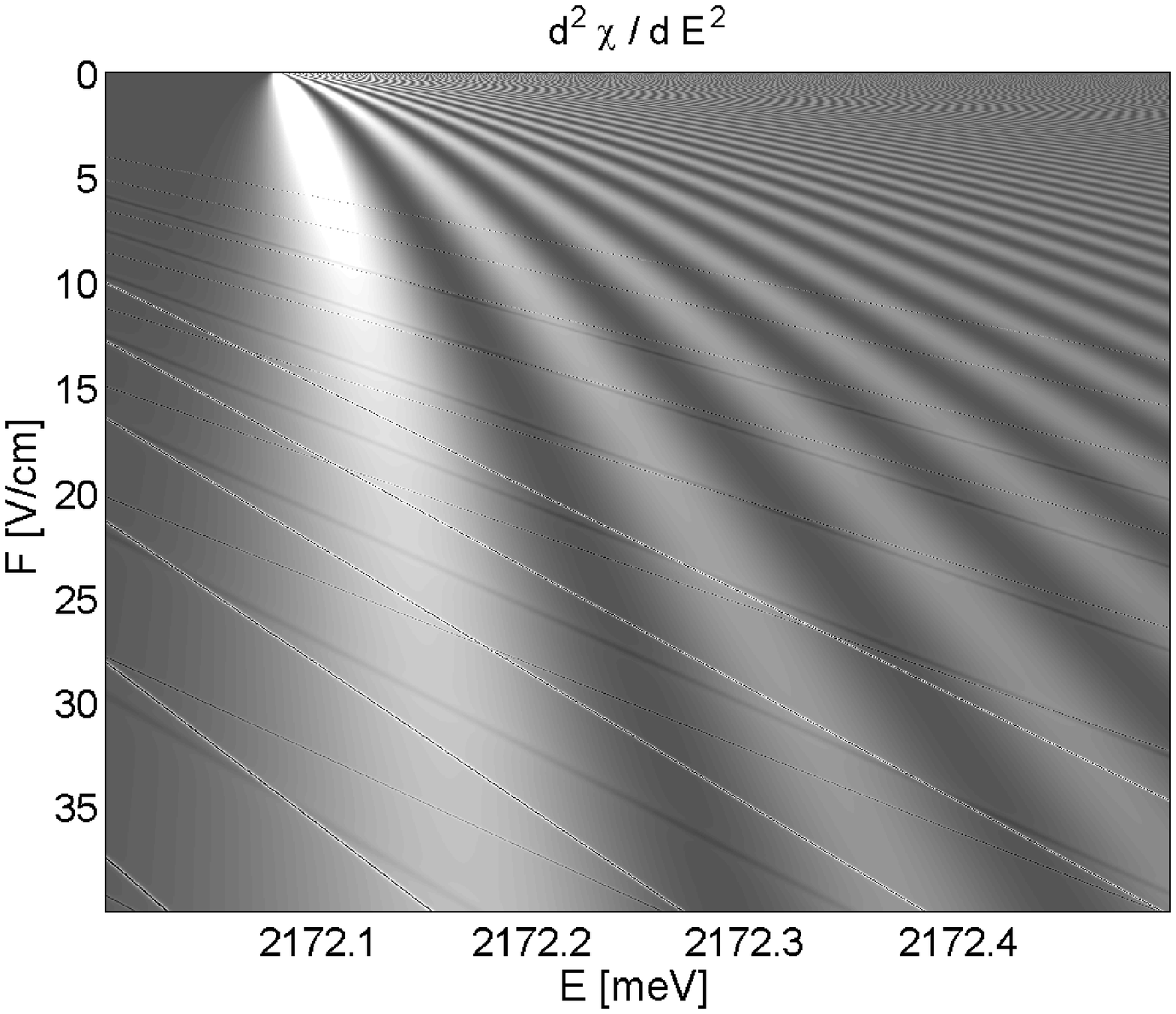}
\caption{\footnotesize The spectrum of the second derivative of susceptibility for a
Cu$_2$O crystal and the applied field strength
\hbox{$F$=0-40 V/cm}. The long period F-K oscillations are superimposed on Stark-shifted excitonic states, seen as thin lines. }\label{Fig5}
\end{figure}

\begin{figure}
\includegraphics[width=0.9\linewidth]{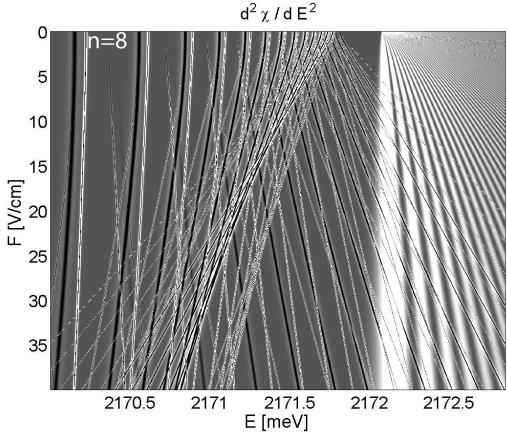}
\caption{\footnotesize The same as in Fig. \ref{Fig5}, over a wider range of energy. The border of F-K oscillations at \hbox{2172 meV} is readily visible.
Excitonic states with $n$=8-20 are taken into account.}\label{Fig6}
\end{figure}

\begin{figure}
\includegraphics[width=0.8\linewidth]{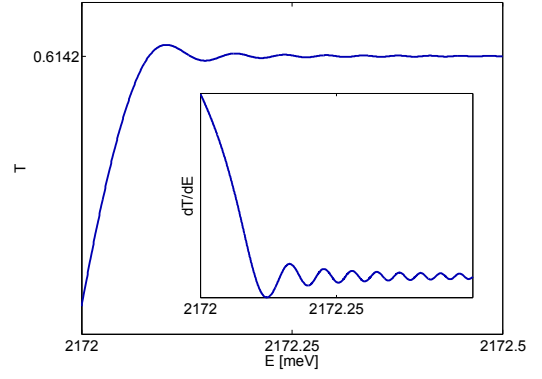}
\caption{\footnotesize Transmission spectrum for a Cu$_2$O crystal
with thickness 30 $\mu$m, calculated by the formula (\ref{transmission}).\vspace{1em}}\label{Fig7}
\end{figure}

\begin{figure}
\includegraphics[width=0.8\linewidth]{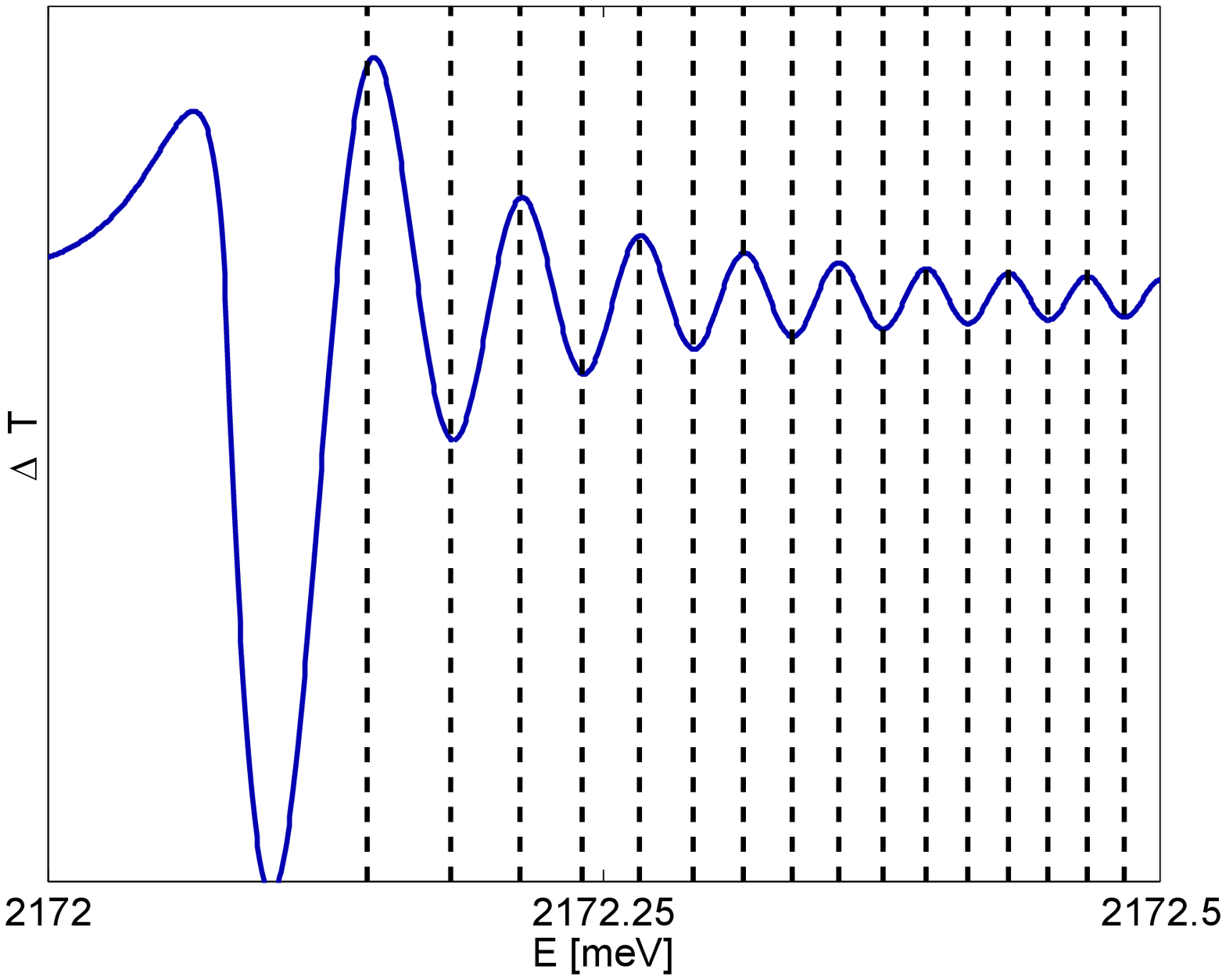}
\caption{\footnotesize The spectrum of the difference $\Delta
T=T(F=10~\mbox{V/cm})-T(F=0)$. The dashed lines indicate the position of peaks
obtained from the Eq. (\ref{FKE}).}\label{Fig8}
\end{figure}

The complete results will contain the exciton effect, which in our
theoretical treatment is related to the resonant denominator $Q$.
The results are displayed in Figs. \ref{Fig4}-\ref{Fig8}. In Fig.
\ref{Fig4} we illustrate the influence of excitons on the shape of the
susceptibility. The impact of exciton manifests in increasing of the absorption;
the FK oscillations increase  and  move towards higher energies while
their period and phase remains almost the same.
 The FK oscillations become more evident when we
plot higher derivatives of the susceptibility.
 It is shown in
Figs. \ref{Fig5}-\ref{Fig6} where we show the dependence of the
second derivative ${\rm d}^2\chi/{\rm d}E^2$ on the excitation
energy. The Fig. \ref{Fig5} shows the imaginary part of susceptibility as a
function of energy and field strength. For clarity, the brightness
is proportional to the second derivative with respect to the
energy. The F-K effect is visible as sinusoidal oscillations with
period proportional to the applied field. One can see that the
first maximum occurs just above the band gap, at 2172 meV. The
thin lines are Stark shifted absorption lines of $P$ and $F$
excitons with principal number up to n=20, calculated according to
our method.\cite{Zielinska.PRB.2016.b} The Fig. \ref{Fig6} shows the full
absorption spectrum in a wide range of energy below and over the
band gap, highlighting both the F-K effect and excitonic states.
One can see that due to the Stark shift, for sufficiently strong
field F, the excitonic maxima overlap with the F-K effect. Having
the susceptibility, we can calculate the optical functions from
the relations (\ref{wzoryfcjeoptyczne}). According to the
discussion presented in Ref.\cite{Zielinska.PRB.2016.b}  the
polaritonic effect in Cu$_2$O can be neglected so optical
functions can be applied with the help of the usual formulas for a
dielectric slab of thickness; the transmissivity $T$ is given by
expression

\begin{eqnarray}\label{transmission}
&&T=\frac{16\vert n\vert^2}{\vert(1+n)^2\vert^2}\;e^{-\alpha
L}=\nonumber\\&&=\frac{16\left(n_1^2+n_2^2\right)}{\left[\left(1+n_1\right)^2-n_2^2\right]^2+4n_2^2\left(1+n_1\right)^2}\;e^{-\alpha
L}.
\end{eqnarray}
Here
\begin{equation}\label{alpha}
\alpha=2\frac{\hbar\omega}{\hbar c}\hbox{Im}\,n
\end{equation}
denotes the absorption coefficient, and $n$ is the complex
refraction coefficient $n=\sqrt{\epsilon_b+\chi}=n_1+in_2$. The
results obtained from Eq. (\ref{transmission}) are displayed in
Figs. \ref{Fig7} and \ref{Fig8}. As can be seen from Fig.
\ref{Fig7} the transmission spectrum has characteristic FK
oscillations in the energy region above the gap. Again as
expected, their amplitude and periodicity relay on the electric
field strength. This effect is more pronounced in Fig. \ref{Fig8},
where the spectrum of transmission difference  as the of energy is
displayed. The positions of oscillations' minima and maxima are in
perfect agreement with peaks for excitation energies predicted by
formula (\ref{FKE}).

\section{Discussion and conclusions}
 We wish to summarize briefly the results we  have obtained by applying the dynamical density matrix approach for
 the optical properties of semiconductor's with Rydberg excitons  exposed to a static electric field.
We have developed a simple mathematical procedure to calculate
electro-optical functions of semiconductor crystal with symmetry
where $P$-exciton transitions are dipole allowed. For excitation
energies larger than the fundamental gap we observe oscillations
in all optical functions which are identified with Franz-Keldysh
oscillations. Their periodicity with respect to the excitation
energy, the amplitudes and the dependence on the applied field
strength was calculated and presented in the form of analytical
expressions.  The results differ from the known results on FK
effect for $S$ excitons.
  We have also  examined the influence  of the coherence of the carriers with the electromagnetic field.
  The presented method has been used to investigate electro-optical functions of Cu$_2$O crystal  for the case of
   normal incidence and the static electric field applied in the same direction.

Franz-Keldysh effect provides the optoelectronic mechanism to
create and  control the electro-modulations which might be an
essential and flexible tool for constructing optical compatible
output devices e.g., a modulator or detectors with an off-chip
laser. The copper oxide-based optoelectronic modulators
employing Franz-Keldysh effect might show great promise in meeting
the strict energy requirements with controlled modification of the reflection/transmission
modulation.

The experimental data  for FK effect in Cu$_2$O are not available
yet, but we hope that our theoretical considerations  might
stimulate  experiments of the electro-optical properties of this
crystal for above gap regime. We conclude that the dynamical
density matrix approach is well suited to describe the macroscopic
fields (static and dynamic) and the microscopic excitons in all
limits of physical interest.




\begin{thebibliography}{99}
\bibitem{Frenkel}
Y. Frenkel, On transformation of light into heat in solids, I,
Phys. Rev. \textbf{37}, 17 (1931); On transformation of light into
heat in solids, II, Phys. Rev. \textbf{37}, 1276 (1931).
\bibitem{Peierls}
R. Peierls, Zur Theorie der Absorptionsspektren fester K\"{o}rper,
Annln. Physk \textbf{13}, 905 (1932). \bibitem{Wannier} G. H.
Wannier, The Structure of Electronic Excitation Levels in
Insulating Crystals, Phys. Rev. \textbf{52}, 191 (1937).
\bibitem{Mott} N. F. Mott, Conduction in polar crystals, II,
Trans. Faraday Soc. \textbf{34}, 500 (1938).
\bibitem{Gross} Gross, E. F., Karryjew, N. A. Dokl. Akad. Nauk
SSSR 84, 471�474 (1952).
 \bibitem{Knox}
  R. S. Knox,\emph{ Theory of Excitons} (Academic Press, New York, 1963).
  \bibitem{Bas85}
F. Bassani and G. Pastori-Parravicini, {\sl Electronic States and
Optical Transitions in Solids} (Pergamon Press, Oxford, 1975).
  \bibitem{Agranovich}
V. M. Agranovich and V. L. Ginzburg, {\sl Crystal Optics with
spatial Dispersion and Excitons} (Springer Verlag, Berlin, 1984).
  \bibitem{StB87}
A. Stahl and I. Balslev, {\sl Electrodynamics of the Semiconductor
Band Edge} (Springer-Verlag, Berlin-Heidelberg-New York, 1987).
\bibitem{LaRocca}
G. La Rocca, \emph{Wannier-Mott Excitons in Semiconductors}. In:
\emph{Electronic Excitations in Organic Based Nanostructures}, ed.
by V. M. Agranovich and G. F. Bassani, Thin Films and
Nanostructures, vol. \textbf{31}, Elsevier, Amsterdam, 2003, pp.
97-128.
\bibitem{Bassani_2003} F.
Bassani,\emph{ Polaritons}. In:  \emph{Electronic Excitations in
Organic Based Nanostructures}, ed. by V. M. Agranovich and G. F.
Bassani, Thin Films and Nanostructures , Vol. \textbf{31}, 129-183
(2003) (Elsevier, Amsterdam, 2003, ISBN: 0-12-533031-6).
\bibitem{Agran2009}
 V. M. Agranovich, \emph{Excitations in Organic Solids} (Oxford
 University Press, Oxford, 2009, ISBN 978 0 19 9234417).
 \bibitem{YuCardona}
 P. Y. Yu and M. Cardona, \emph{Fundamentals of Semiconductors}, 4$^{\rm
 th}$ Ed., (Springer, Berlin-Heidelberg,
 ISBN 978-3-642-00709-5, 2010).
 \bibitem{Klingshirn}
C. F. Klingshirn, \emph{Semiconductor Optics}, 4$^{\rm th}$ ed.
(Springer, Berlin Heidelberg, ISBN 978-3-642-28362-8, 2012).
\bibitem{Froelich} D. Fr\"{o}hlich, A. Kulik, B. Uebbing, A.
Mysyrowicz, V. Langer, H. Stolz, and W. von der Osten, Coherent
Propagation and Quantum Beats of Quadrupole Polaritons in Cu$_2$0,
Phys. Rev. Lett.  67, 2343 (1991).
\bibitem{Nikitine_1959}
S. Nikitine, Experimental investigations of exciton spectra in
ionic crystals, Philisophical Magazine \textbf{4}, 1 (1959).
\bibitem{Froelich1982}
D. Fr\"{o}hlich and R. Kenklies, Polarization dependence of
two-photon magnetoabsorption of the 1s exciton in Cu$_2$O, Phys.
Stat. Sol. (b) \textbf{111}, 247 (1982).
\bibitem{Ghijsen}
J. Ghijsen, L. H. Tjeng, J. van Elp, H. Eskes, J. Westerink, G. A.
Sawatzky, and  M. T. Czyzyk, Electronic structure of Cu$_2$O and
CuO, Phys. Rev. B. \textbf{38}, 11 322 (1988).

\bibitem{Jolk}
A. Jolk, M. J\"{o}rger, and C. Klingshirn, Exciton lifetime, Auger
recombination, and exciton transport by calibrated differential
absorption spectroscopy in Cu$_2$O, Phys. Rev. B. \textbf{65},
245209 (2002).
\bibitem{Joerger_2005}
M. J\"{o}rger, T. Fleck, C. Klingshirn, and  R. von Baltz,
Midinfrared properties of cuprous oxide: High-order lattice
vibrations and intraexcitonic transitions of the 1s paraexciton,
 Phys. Rev. B. \textbf{71}, 235210 (2005).
 \bibitem{Stolz_2012}
 H. Stolz, R. Schwartz, F. Kieseling,
S. Som, M. Kaupsch, S. Sobkowiak, D. Semkat, N. Naka,  T. Koch,
and H. Fehske, Condensation of excitons in Cu$_2$O at ultracold
temperatures: experiment and theory, New Journal of Physics
\textbf{14},  105007 (2012).
\bibitem {Kazimierczuk}
T. Kazimierczuk, D. Fr\"{o}hlich, S. Scheel, H. Stolz, and M.
Bayer, Giant Rydberg excitons in the copper oxide Cu$_2$O, Nature
\textbf{514}, 344 (2014).

\bibitem{Thewes}
J. Thewes, J. Heck\"{o}tter, T. Kazimierczuk, M. A{\ss}mann, D.
Fr\"{o}hlich,  M. Bayer, M. A. Semina, and M. M. Glazov,
Observation of High Angular Momentum Excitons in Cuprous Oxide,
Phys. Rev. Lett. \textbf{115}, 027402 (2015).

\bibitem{Zielinska.PRB}
S. Zieli\'{n}ska-Raczy\'{n}ska, G. Czajkowski, and D. Ziemkiewicz,
Optical properties of Rydberg excitons and polaritons, Phys. Rev.
B \textbf{93}, 075206 (2016).
 \bibitem{Schoene}
  F. Sch\"{o}ne, S.-O. Kr\"{u}ger, P.
Gr\"{u}nwald, H. Stolz,  M. A\ss mann, J. Heck\"{o}tter, J.
Thewes, D. Fr\"{o}hlich, and M. Bayer, Deviations of the exciton
level spectrum in Cu2O from the hydrogen series, Phys. Rev. B
\textbf{93}, 075203 (2016).
\bibitem{Schweiner} F. Schweiner, J. Main,
and G. Wunner, Linewidths in excitonic absorption spectra of
cuprous oxide, Phys. Rev. B \textbf{93}, 085203 (2016).
\bibitem{Gruenewald}
P. Gr\"{u}nwald, M. A{\ss}mann, J. Heck\"{o}tter, D. Fr\"{o}hlich,
M. Bayer,  H. Stolz, and S. Scheel, Signatures of Quantum
Coherences in Rydberg Excitons, Phys. Rev. Lett. \textbf{117},
133003 (2016).
\bibitem{Hecktoeter_2017} J. Heck\"{o}tter,  M. Freitag, D. Fr\"{o}hlich, M.
A{\ss}mann, M. Bayer, M. A. Semina, and M. M. Glazov, Scaling laws
of Rydberg excitons, Phys. Rev. B \textbf{96}, 125142 (2017).
\bibitem{Schweiner_2017a}
F. Schweiner, J. Main,  G. Wunner, and Ch. Uihlein, Even exciton
series in Cu$_2$O, Phys. Rev. B \textbf{95}, 195201 (2017).
\bibitem{Schweiner_polariton_2017}
F. Schweiner, J. Ertl, J. Main, G. Wunner, and Ch. Uihlein,
Exciton-polaritons in cuprous oxide: Theory and comparison with
experiment, Phys. Rev. B  \textbf{96}, 245202 (2017).
\bibitem{Walther}
V. Walther, R. Johne, and T. Pohl, Giant optical nonlinearities
from Rydberg-excitons in semiconductor microcavities, threearXiv:
1711.01601v1 [cond-mat.quant-gas] 5 Nov 2017.
\bibitem{Zielinska.PRB.2016.b}
S. Zieli\'{n}ska-Raczy\'{n}ska, D. Ziemkiewicz, and G. Czajkowski,
Electrooptical properties of Rydberg excitons, Phys. Rev. B
\textbf{94}, 045205 (2016).
\bibitem{Schweiner.Magnetoexcitons}
 F. Schweiner, J. Main,
G. Wunner, M. Freitag, J. Heck\"{o}tter, Ch. Uihlein, M.
A{\ss}mann, D. Fr\"{o}hlich, and M. Bayer, Magnetoexcitons in
cuprous oxide, Phys. Rev. B \textbf{95}, 035202 (2017).
\bibitem{Zielinska.PRB.2016.c} S.
Zieli\'{n}ska-Raczy\'{n}ska, D. Ziemkiewicz, and G. Czajkowski,
Magneto-optical properties of Rydberg excitons: Center-of-mass
quantization approach, Phys. Rev. B \textbf{95}, 075204 (2017).
\bibitem{Assmann_symmetry}
 M.
A{\ss}mann, J. Thewes, and  M. Bayer, Quantum chaos and breaking
of all antiunitary symmetries in Rydberg excitons, Nature
Materials \textbf{15}, 741 (2016).
\bibitem{Schweiner_Rommel}
F. Schweiner, P. Rommel, J. Main, and G. Wunner, Exciton-phonon
interaction breaking all antiunitary symmetries in external
magnetic fields, Phys. Rev. B \textbf{96}, 035207 (2017).

\bibitem{Schweiner_Symmetry}
F. Schweiner, J. Main,  and G. Wunner, Magnetoexcitons Break
Antiunitary Symmetries, Phys. Rev. Lett. \textbf{118}, 046401
(2017).
\bibitem{kitamura}
T. Kitamura, M. Takahata1, and N. Naka, Quantum number dependence
of the photoluminescence broadening of excitonic Rydberg states in
cuprous oxide, J. Luminescence \textbf{192}, 808 (2017).
\bibitem{Hecktoetter_2017}
 J. Heck\"{o}tter, M. Freitag, D. Fr\"{o}hlich, M. A{\ss}mann, M.
Bayer, P. Gr\"{u}nwald, F. Sch\"{o}ne, D. Semkat, H. Stolz, and S.
Scheel, Rydberg excitons in the presence of an ultralow-density
electron-hole plasma, arXiv: 1709.00891v1 [cond-mat.mtrl-sci] 4
Sep 2017.

\bibitem{Franz}
W. Franz, Einflu{\ss} eines elektrischen Feldes auf eine optische
Absorptionskante, Z. Naturforschung \textbf{13}a,  484 (1958).
\bibitem{Keldysh}
L. V. Keldysh, Behaviour of Non-Metallic Crystals in Strong
Electric Fields,{\sl Zhurn. Eksp. Teoret. Fiz}. {\bf 34},1138
(1958); (English trans.: {\sl Sov. Phys. JETP} {\bf 7}, 788
(1958)).
\bibitem{Tharmalingam}
K. Tharmalingam Optical Absorption in the Presence of a Uniform
Field, Phys. Rev. \textbf{130}, 2204 (1963).
\bibitem{callaway}
 K. S. Viswanathan and J. Callaway, Dielectric Constant of a Semiconductor in an External Electric
 Field,
Phys. Rev. \textbf{143}, 564 (1966).
\bibitem{Aspnes66}
D. A. Aspnes, Electric-Field Effects on Optical Absorption near
Thresholds in Solids, {Phys. Rev}. {\bf 147}, 554 (1966).
\bibitem{Aymerich}
F. Aymerich and F. Bassani, Electric-field effects on interband
transitions, {Il Nuovo Cimento} \textbf{48} B, 358 (1967).
\bibitem{Ralph}
H. I. Ralph, On the theory of Franz-Keldysh effect, J. Phys. C
\textbf{1}, 378 (1968).
\bibitem{Cardona}
M. Cardona, {\sl Modulation Spectroscopy}, Suppl.11 of {\sl Solid
State Physics}, ed. by F. Seitz, D. Turnbull, and H. Ehrenreich,
(Academic Press, New York-London 1969).
\bibitem {Blos70}
D. F. Blossey, Wannier Exciton in an Electric Field. I. Optical
Absorption by Bound and Continuum States, {Phys. Rev. B} {\bf 2},
3976 (1970).
\bibitem {Blos71}
D. F. Blossey, Wannier Exciton in an Electric Field. II.
Electroabsorption in Direct-Band-Gap Solids, {Phys. Rev. B} {\bf
3}, 1382 (1971).

\bibitem{Lederman}
F. L. Lederman and J. D. Dow, Theory of electroabsorption by
anisotropic and layered semiconductors. I. Two-dimensional
excitons in a uniform electric field, { Phys. Rev. B} {\bf 13},
1633 (1976).

\bibitem{Aronov}
A. G. Aronov and A. S. Ioselevich, {\sl Exciton Electrooptics}.
In: {\sl Excitons}, ed. by E. I. Rashba and M. D. Sturge (North
Holland, Amsterdam 1982).
\bibitem{Schneider}
H. Schneider, A. Fischer, and K. Ploog, Franz-Keldysh oscillations
and Wannier-Stark localization in GaAs/AlAs superlattices with
single-monolayer AlAs barriers, Phys. Rev. B \textbf{45}, 6329
(1992).
\bibitem{Schlichterle}
B. Schlichterle, G. Weiser, M. Klenk, F.Mollot, and Ch.Starck,
Effective masses in In$_{1-x}$Ga$_x$ As superlattices derived from
Franz-Keldysh oscillations, Phys. Rev. B {\bf 52}, 9003 (1995).
\bibitem{Nakayama}
 M.
Nakayama, T. Nakanishi, K. Okajima, M. Ando and H. Nishimura,
Miniband structures and effective masses of GaAs/AlAs
superlattices with ultra-thin layers, Solid State Comm.
\textbf{102}, 803 (1997).
\bibitem{Shen}
 H. Shen and
M. Dutta,  Franz-Keldysh oscillations in modulation spectroscopy,
J. Appl. Phys. \textbf{78}, 2151 (1995); doi: 10.1063/1.360131.
\bibitem{Dressler}
G. Czajkowski, M. Dressler, and F. Bassani, Electro-optical
properties of semiconductor superlattices in the regime of
Franz-Keldysh oscillations, Phys. Rev. B \textbf{55}, 5243 (1997).
\bibitem{Festschrift}
 G. {Czajkowski} and L. {Silvestri}, Electric and magnetic field effects on optical properties of excitons in semiconductor nanostructures. In: {\sl
Electrons and photons in solids - A volume in honour of Franco
Bassani}, edited by  G.~{ Grosso}, G.~C.~{La} {Rocca} and M.~{
Tosi} (Scuola Normale Superiore - Pubblicazioni della classe di
scienze,~Pisa, Italy, 2001), pp.~271-288.
\bibitem{RivistaGC}
G.~Czajkowski, F.~Bassani, and L.~Silvestri, Excitonic Optical
Properties of Nanostructures: Real Density Matrix Approach,
{Rivista del Nuovo Cimento} \textbf{26}, 1-150 (2003).
\bibitem{Dressler_EPJ}
M. Dressler, F. Bassani, and G. Czajkowski, Electro-optical properties of excitons in polydiacetylene
chains, Eur. Phys. J. B \textbf{10}, 681 (1999).
\bibitem{Brazil}
 J. R. Madureira, M. Z. Maialle,  M. H. Degani, Franz-Keldysh Effect in Semiconductor T-Wire
in Applied Magnetic Field, Brazilian Journ. Phys.
 \textbf{34}, 663 (2004).
 \bibitem{Schaevitz}
 R. K. Schaevitz, D. S. Ly-Gagnon, J. E. Roth, E. H. Edwards, and
 D. A. B. Miller, Indirect absorption in germanium quantum wells,
 AIP Advances \textbf{1}, 032164 (2011).
 \bibitem{Lee}
 S. J. Lee,
Ch. W. Sohn, H.-J. Jo, I. S. Han, J. S. Kim, S. K. Noh, H. Choi,
J.-Y. Leem, Temperature Dependence of the Photovoltage from
Franz-Keldysh Oscillations in a GaAs p+-i-n+ Structure, J. Korean
Phys. Society \textbf{67}, 916, (2015).
\bibitem{Patent}
Patent US 5365334 A, Micro photoreflectance semiconductor wafer
analyzer; Photoreflectance spectral analysis of semiconductor
laser structures US 6195166 B1; Optical measuring method for
semiconductor multiple layer structures and apparatus therefor US
7038768 B2.
\bibitem{Abram81}
{\sl Handbook of Mathematical Functions}, edited by M. Abramowitz
and I. Stegun (Dover Publications, New York 1965).
\bibitem{Analysis}
E. T. Whittaker and G. N. Watson,\emph{ A Course of Modern
Analysis} (Cambridge Un. Press, Cambridge, 1935, Reissued in the
Cambridhe Math. Library Series, 1996, ISBN 05215880703).

\bibitem{Grad}
I. S. Gradshteyn and I. M. Ryzhik, \emph{Table of Integrals,
Series, and Products}, edited by A. Jeffrey and D. Zwillinger, 7th
Edition (Academic Press, Elsevier, Amsterdam, 2007,
ISBN-13:978-0-12-373637-6).

\bibitem{Airy}
O. Vall\'{e}e and M. Soares, \emph{Airy Functions and Applications
to Physics} (Imperial College Press, London, 2004, ISBN:
1-86094-478-7).
\end{thebibliography}
\end{document}